**Broadening of In-Field Superconducting Transitions in Hydrides**

***Comment on "Nonstandard Superconductivity or No Superconductivity in Hydrides under High Pressure"*** [J. E. Hirsch and F. Marsiglio in *Phys. Rev. B* **103**, 134505 (2021)]

E. F. Talantsev[1]*, V. S. Minkov[2], F. F. Balakirev[3], M. I. Eremets[2]*

[1]*M. N. Mikheev Institute of Metal Physics, S. Kovalevskoy St. 18, Ekaterinburg 620108, Russian Federation*
[2]*Max Planck Institute for Chemistry; Hahn-Meitner-Weg 1, Mainz 55128, Germany*
[3]*NHMFL, Los Alamos National Laboratory, MS E536, Los Alamos, New Mexico 87545, USA*

*J. E. Hirsch and F. Marsiglio in their publication, Phys. Rev. B* ***103****, 134505 (2021), assert that hydrogen-rich compounds do not exhibit superconductivity. Their argument hinges on the absence of broadening of superconducting transitions in applied magnetic fields. We argue, that this assertion is incorrect, as it relies on a flawed analysis and a selective and inaccurate report of published data, where data supporting the authors´ perspective are highlighted while data demonstrating clear broadening are disregarded.*

In the original paper[1], the authors claim the absence of broadening of resistive superconducting transitions in hydrides under applied magnetic fields. This absence of broadening serves as their primary argument against the existence of superconductivity in hydrides, as broadening is typically observed in most well-known superconductors at ambient pressure[1]. Broadening arises due to the penetration of magnetic vortices in type-II superconductors, leading to vortex motion associated with electrical currents and subsequent dissipation. Consequently, the width of the resistive transition, $\Delta T$, is expected to increase with the magnetic field as $\sim H^{2/3}$ relationship, as described by Tinkham[2]. However, it is important to note that this is not a universal law, as the degree of broadening depends on various factors, including vortex pinning mechanisms and the particularities of the superconducting order parameter fields[3].

For example, in cuprates, the presence of a vortex liquid phase results in a substantial broadening of the superconducting transition under external magnetic fields[3], while iron-based superconductors exhibit relatively smaller broadening[4]. There have been reports of even the narrowing of the transition width in $MgB_2$ under applied magnetic fields[5]. Therefore, the absence or the magnitude of broadening should not be taken as definite evidence that a material is not a superconductor.

Moreover, available data indicate broadening in a variety of hydrides. A comprehensive analysis of the in-field broadening of superconducting transitions in hydrides was conducted in Ref.[6]. However, J. E. Hirsch and F. Marsiglio ignored these available data, instead choosing to emphasize data that supports their hypothesis (as shown in Figure 1). Based on this inaccurate report, the authors have made an overgeneralization that all hydride superconductors are not, in fact, superconductors. Alarmingly, the authors continue to persistently employ this approach in their subsequent works[7-9].

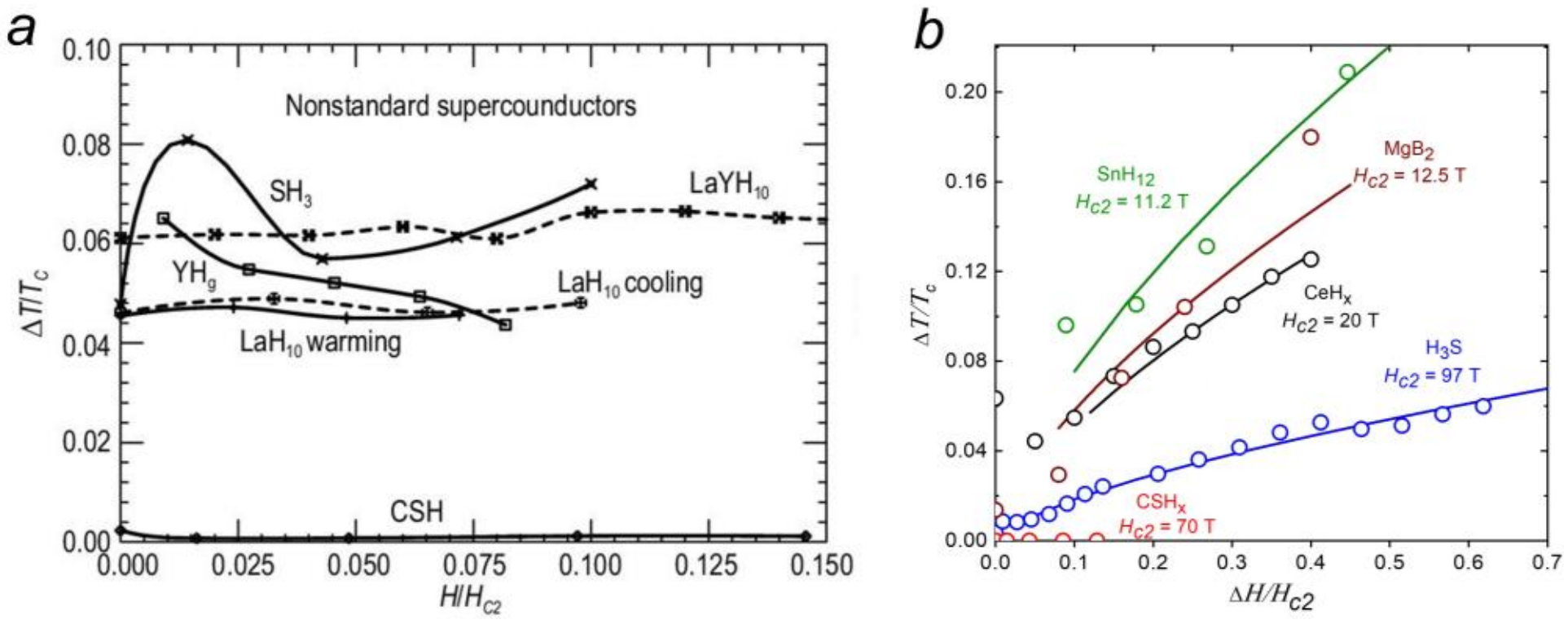

**Figure 1.** Broadening of the superconducting transition under external magnetic fields in hydrogen-rich compounds at high pressures. (a) Plot reproduced from Ref.[1] only presents the data selectively chosen to support the authors' claim of the absence of broadening of the superconducting transition. (b) Broadening of the superconducting transition in various hydrogen-rich compounds at high pressures, including $SnH_{12}$[10], $CeH_x$[11], $H_3S$[12], and C-S-H[13] in comparison with non-hydride $MgB_2$ ambient-pressure superconductor[14], reproduced from Ref.[15]. The selected coordinates facilitate easy comparison. The solid lines serve as visual aids for the expected broadening, $\Delta T(B) \sim B^{2/3}$, at high magnetic fields[2].

**In-depth analysis of the magnetic field-induced broadening of superconducting transitions.**

Reiterating central message expressed in the main text[1], we argue that highly-compressed hydrogen-rich superconductors do not exhibit a different in-field resistive transition when compared to other classes of superconductors. This perspective contradicts the findings of the authors[1], who reported that high-pressure hydrides exhibit a transition width that *"either stays the same as for H=0 or even decreases with applied field"*. Based on their observations, the authors[1] concluded that compressed hydrides are either nonstandard superconductors or not superconductors at all.

In this work, we provide the in-depth analysis of available experimental data reported by various research groups. This data includes, but is not limited to: (a) ambient-pressure and (b) high-pressure superconductors, in which the resistive transition narrows with applied magnetic field; and (c) highly-pressurized hydrides, in which the resistive transition broadens with applied magnetic fields. Our analysis demonstrates that the claim made by the authors[1] is incorrect.

Based on our analysis, we conclude that the authors[1] cherry-picked data in support of their claim while excluding datasets that contradict it. We argue that the assertion of "*nonstandard superconductivity or no superconductivity in hydrides*" arises from a bias in data selection.

Additionally, we highlight that the authors[1] failed to account for the existence of the intermediate state in the magnetic phase diagram of type-I superconductors. Therefore, we call for further scrutiny of recent papers by the authors[7, 9, 16-18] to reevaluate the correctness of their interpretation of magnetic properties of highly-compressed hydrides.

**I. Detailed explanation of the primary message of the Comment.**

The authors[1] claimed that highly-compressed hydrides exhibit a narrowing of the resistive transition width $\Delta T_c\left(B_{appl}\right)$ in an applied magnetic field, $B_{appl}$, which has been never observed in "*standard conventional and unconventional type-II superconductors*". This is because the transition "*broadening either stays the same as for H=0 or even decreases with applied field*" in hydrides at high pressures[1]. Based on this, the authors[1] concluded that "*the signals interpreted as superconductivity*" measured in hydrides *"are either experimental artifacts or they signal other interesting physics but not superconductivity"*.

Here, we demonstrate the following:

1. There is no universal in-field transition width $\Delta T_c\left(B_{appl}\right)$ dependence that can serve as the criterion for determining the existence of the superconductivity in any given material;
2. Hydrides at high pressures exhibit a similar in-field transition width $\Delta T_c\left(B_{appl}\right)$ dependence to other superconductors, such as $MgB_2$;
3. The authors[1] employed undisclosed method(s) for extracting the transition width $\Delta T_c\left(B_{appl}\right)$ from raw experimental $R\left(T, B_{appl}\right)$ data;
4. The authors[1] selectively presented datasets that support their claims while excluding datasets that contradict them;
5. The authors[1] misinterpreted the magnetic phase diagrams of superconductors by overlooking the existence of the intermediate state in type-I superconductors.

Leaving aside all the extended narratives expressed by the authors in Ref.[1], here, we aim to demonstrate that the primary claim of the authors:

*"In standard conventional and unconventional type-II superconductors it is empirically observed that the resistive transition is broadened in an applied magnetic field, the more so the larger the field... This is the expected behavior in all standard type-II superconductors, which can be understood theoretically using concepts developed more than 50 years ago that apply to both conventional and unconventional superconductors*[2]*... We conclude that if hydrides under pressure are truly*

*superconductors, they do not obey the same physical principles that all other type-II superconductors, whether conventional or unconventional, obey."*

is incorrect because there is no universal resistive transition width broadening trend *vs.* applied magnetic field in type-I and type-II superconductors. Both in-field broadening and narrowing trends of the resistive transition have been observed in experiments, and both trends have been observed for the same samples in applied magnetic field of different ranges.

## II. The transition width definition.

As we mentioned earlier, the authors[1] did not provide details for determining the transition width, $\Delta T_c(B_{appl})$, which they utilized to extract this value from experimental $R(T, B_{appl})$ data.

Furthermore, the authors[1] did not include any plots displaying the extracted $\Delta T_c(B_{appl})$ alongside the $R(T, B_{appl})$ curve from which the width was obtained.

Despite our repeated requests to the authors[1] for clarification on their criteria and for providing the extracted $\Delta T_c(B_{appl})$ in conjunction with the $R(T, B_{appl})$ curves, we received no response.

Given the unknown transition width definition used in Ref.[1], this work employs a widely accepted definition for the resistive transition width[5, 19, 20]:

$$\frac{\Delta T_{c,0.9-0.1}(B_{appl})}{T_c(B_{appl}=0)} = \frac{T_{c,0.9}(B_{appl}) - T_{c,0.1}(B_{appl})}{\left(\frac{T_{c,0.9}(B_{appl}=0)+T_{c,0.1}(B_{appl}=0)}{2}\right)} \quad (1)$$

where $T_{c,0.9}(B_{appl}) = \frac{R(T,B_{appl})}{R(T_{c,onset},B_{appl})} = 0.90$ and $T_{c,0.1}(B_{appl}) = \frac{R(T,B_{appl})}{R(T_{c,onset},B_{appl})} = 0.10$.

In the following sections, we employ the same plot style as in Ref.[1], featuring the horizontal $\frac{\Delta T_c(B_{appl})}{T_c(B_{appl}=0)}$ axis and the vertical $\frac{B_{appl}}{B_{c2}(0)}$ axis (where $B_{c2}(0)$ is the ground state upper critical field).

## III. Type-I superconductors.

Before reporting the results of our analysis in type-II superconductors, we present experimental data for $\frac{\Delta T_c(B_{appl})}{T_c(B_{appl}=0)}$ *vs.* $\frac{B_{appl}}{B_{c,therm}(0)}$ (where $B_{c,therm}(0)$ is ground state thermodynamic critical field) for zinc[21] and cadmium[22] in Figure 2. We use a linear-log plot in the style introduced by Schooley[22]. It is evident that type-I superconductors also exhibit trends for $\frac{\Delta T_c(B_{appl})}{T_c(B_{appl}=0)}$ *vs.* $\frac{B_{appl}}{B_{c,therm}(0)}$ dependences.

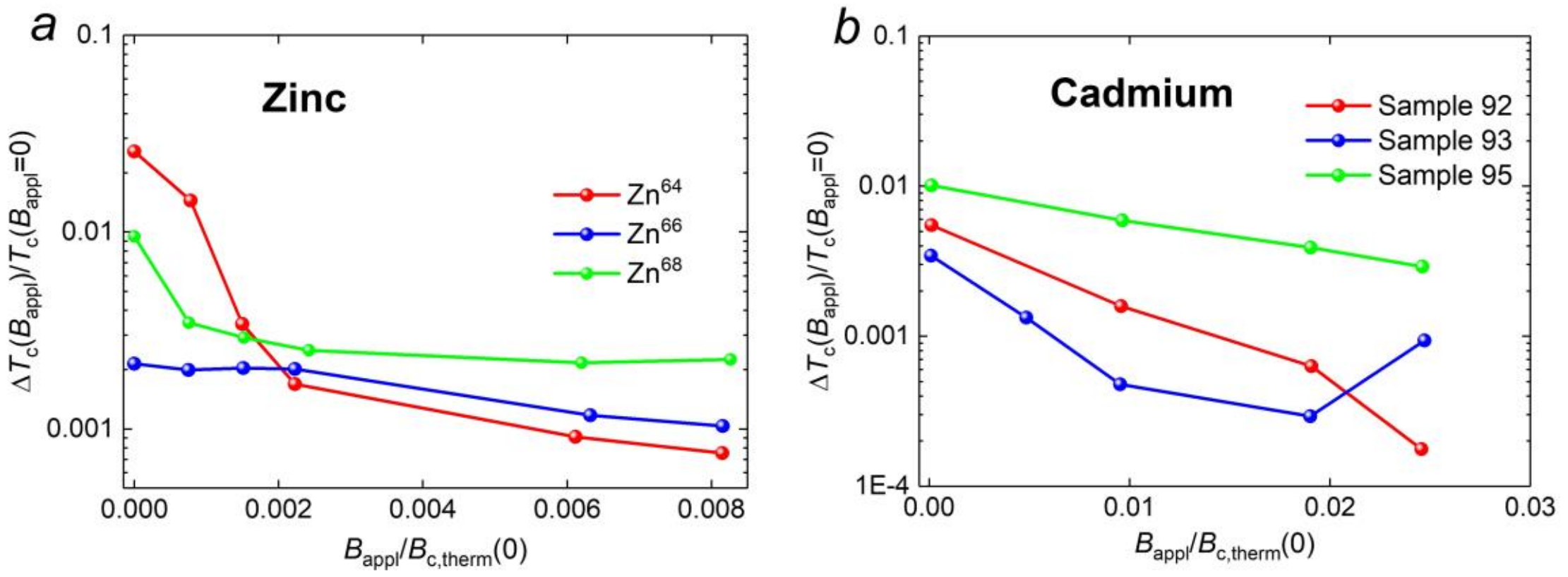

**Figure 2.** Dependences of transition width, $\frac{\Delta T_c(B_{appl})}{T_c(B_{appl}=0)}$, *vs.* applied magnetic field, $\frac{B_{appl}}{B_{c,therm}(0)}$, in type-I superconductors. (a) Experimental data for zinc isotopes reported by Fassnacht and Dillinger[21], with the following values used: $Zn^{64}$ ($B_{c,therm}(0) = 5.394\ mT$ and $T_c(B_{appl} = 0) = 0.8553\ K$); $Zn^{66}$ ($B_{c,therm}(0) = 5.377\ mT$ and $T_c(B_{appl} = 0) = 0.8456\ K$); $Zn^{68}$ ($B_{c,therm}(0) = 5.313\ mT$ and $T_c(B_{appl} = 0) = 0.8364\ K$). (b) Experimental data for cadmium, as reported by Schooley[22], with the following values used: $B_{c,therm}(0) = 2.73\ mT$ and $T_c(B_{appl} = 0) = 0.515\ K$.

However, at some high $\frac{B_{appl}}{B_{c,therm}(0)}$ values (e.g., sample 93 of cadmium[22]), the narrowing trend changes into a broadening one. A theoretical explanation for the effect of transition width narrowing in an applied magnetic field has been proposed[23].

## IV. Type-II superconductors (non-hydrides).

In their Figure 11a, the authors[1] presented the $\frac{\Delta T_c(B_{appl})}{T_c(B_{appl}=0)}$ *vs.* $\frac{B_{appl}}{B_{c2}(0)}$ relationships for several representative compounds from the main families of non-hydride superconductors: $YBa_2Cu_3O_{7-x}$, $MgB_2$, $Ba(Fe_{0.645}Ru_{0.355})_2As_2$, NbN, $Nb_3Sn$, $K_3C_{60}$. All the $\frac{\Delta T_c(B_{appl})}{T_c(B_{appl}=0)}$ *vs.* $\frac{B_{appl}}{B_{c2}(0)}$ relationships presented in Figure 11a in Ref.[1] exhibited the broadening trend of transition width at applied magnetic fields.

In Figure 3, we demonstrate the $\frac{\Delta T_c(B_{appl})}{T_c(B_{appl}=0)}$ *vs.* $\frac{B_{appl}}{B_{c2}(0)}$ relationships for $MgB_2$, $Na(Fe_{0.99}Co_{0.01})As$, and $Bi_2Sr_2CaCu_2O_8$, along with the relationships reported in Figure 11a in Ref.[1] (all shown in black in all panels for comparison). These materials are all type-II superconductors and exhibit a narrowing trend in the transition width *vs.* applied magnetic field. Detailed analyses of $R(T, B_{appl})$ datasets for $MgB_2$ are provided in *Appendix A* in Supplementary Materials, for $Na(Fe_{0.99}Co_{0.01})As$ in *Appendix B* in Supplementary Materials, and for $Bi_2Sr_2CaCu_2O_8$ in *Appendix C* in Supplementary Materials.

Based on data presented in Figure 3, one can conclude that the universal broadening trend of $\frac{\Delta T_c(B_{appl})}{T_c(B_{appl}=0)}$ *vs.* $\frac{B_{appl}}{B_{c2}(0)}$ claimed in Ref.[1], which states, "*empirically observed... in all standard type-II superconductors, which can be understood theoretically using concepts developed more than 50 years ago that apply to both conventional and unconventional superconductors[2]*" is incorrect because experimental data in Figure 3 show the opposite.

Another important issue is that there is no universal $\frac{\Delta T_c(B_{appl})}{T_c(B_{appl}=0)}$ *vs.* $\frac{B_{appl}}{B_{c2}(0)}$ relationship for $MgB_2$. Instead, the *fabrication technique and sintering procedure*, rather than the nominal *sample composition*, determine the $\frac{\Delta T_c(B_{appl})}{T_c(B_{appl}=0)}$ *vs.* $\frac{B_{appl}}{B_{c2}(0)}$ trend. This is demonstrated in Figure 3b, where we present $\frac{\Delta T_c(B_{appl})}{T_c(B_{appl}=0)}$ *vs.* $\frac{B_{appl}}{B_{c2}(0)}$ data reported by Wang *et al.*[5] for three different $MgB_2$+SiC samples (details can be found in Ref.[5]).

Therefore, we come to the central issue: the claim made by the authors[1], that the $\frac{\Delta T_c(B_{appl})}{T_c(B_{appl}=0)}$ *vs.* $\frac{B_{appl}}{B_{c2}(0)}$ *relationship is linked exclusively to sample composition, is fundamentally incorrect*.

Considering that all near-room temperature superconductors are synthesized and annealed under harsh conditions (e.g., at pressure of 100-250 GPa and with pulsed temperatures rising up to 3000 K), it is not unexpected that local atomic disorder, composition and pressure variations across the sample area can vary. Thus, the $\frac{\Delta T_c(B_{appl})}{T_c(B_{appl}=0)}$ *vs.* $\frac{B_{appl}}{B_{c2}(0)}$ relationship can be different for different hydride samples, even if the nominally formed phase and global sample compositions are the same.

It is also important to note that the $\frac{\Delta T_c(B_{appl})}{T_c(B_{appl}=0)}$ *vs.* $\frac{B_{appl}}{B_{c2}(0)}$ relationship are not monotonic (see Figure 3). This applies equally to $MgB_2$, $Na(Fe_{0.99}Co_{0.01})As$, and $Bi_2Sr_2CaCu_2O_8$ superconductors.

In conclusion, based on the above, we argue that the $\frac{\Delta T_c(B_{appl})}{T_c(B_{appl}=0)}$ *vs.* $\frac{B_{appl}}{B_{c2}(0)}$ relationship is not the fundamental dependence and cannot serve as the criterion for the existence or absence of the superconducting state in any material.

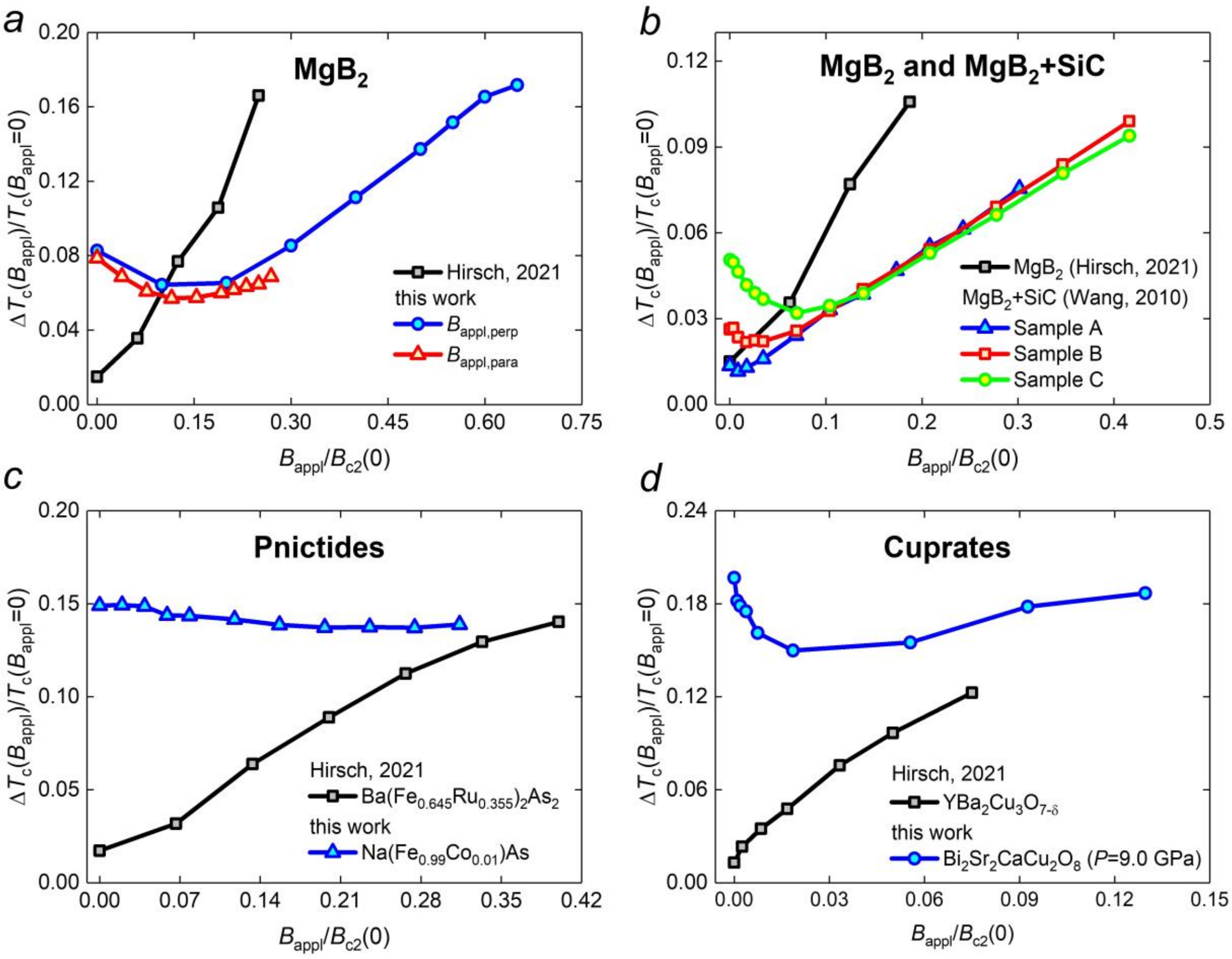

**Figure 3.** Transition width, $\frac{\Delta T_c(B_{appl})}{T_c(B_{appl}=0)}$, *vs.* applied magnetic field, $\frac{B_{appl}}{B_{c2}(0)}$, for various type-II superconductors. (a) $MgB_2$: black squares from Ref.[1] extracted from experimental data reported in Ref.[24]; blue circles deduced in this work by applying Equation 1 to data reported in Ref.[25] for magnetic fields applied perpendicular to the (0001) plane of the film (details are in *Appendix A* in Supplementary Materials); red triangles deduced in this work by applying Equation 1 to data reported in Ref.[25] for magnetic fields applied parallel to the (0001) plane of the film (details are in *Appendix A* in Supplementary Materials). (b) doped $MgB_2$: black squares from Ref.[1] extracted from experimental data reported in Ref.[24]; data for $MgB_2$ doped with SiC reported in Ref.[5] ($T_c(B_{appl}=0)=39.2\,K$ and $B_{c2}(0)=28.9\,T$ were used): sample A (blue triangles) prepared by a diffusion reaction method sintered at 750 °C for 10 h; sample B (red squares) prepared by a diffusion reaction method sintered at 850 °C for 10 h; sample C (green circles) prepared by an *in situ* reaction (mixed) method with sintering at 850 °C for 10 h. (c) Pnictides: data for $Ba(Fe_{0.645}Ru_{0.355})_2As_2$ from Ref.[1] (black squares) extracted from experimental data reported in Ref.[26]; data for $Na(Fe_{0.99}Co_{0.01})As$ (blue triangles) deduced in this work by applying Equation 1 to data reported in Ref.[27] (details are in *Appendix B* in Supplementary Materials). (d) Cuprates: data for $YBa_2Cu_3O_{7-\delta}$ from Ref.[1] (black squares) extracted from experimental data reported in Ref.[2]; data for $Bi_2Sr_2CaCu_2O_8$ (blue circles) deduced in this work by applying Equation 1 to data reported in Ref.[28] (details are in *Appendix C* in Supplementary Materials).

## V. Highly-compressed scandium.

The detailed analysis of the $\frac{\Delta T_c(B_{appl})}{T_c(B_{appl}=0)}$ *vs.* $\frac{B_{appl}}{B_{c2}(0)}$ relationships for scandium under high pressures, exhibiting record high-$T_c$ values for elements[29, 30], is presented in *Appendix D* in Supplementary Materials. Ground state upper critical field values, $B_{c2}(0,P)$, were determined and depicted in Figure A9 in *Appendix D* in Supplementary Materials. These values were then used to calculate data for the $\frac{\Delta T_c(B_{appl},P)}{T_c(B_{appl}=0,P)}$ *vs.* $\frac{B_{appl}}{B_{c2}(0,P)}$ plots, as shown in Figure 4. While an overall trend of transition width broadening *vs.* applied field can be observed, it is important to note that each curve has anomalies associated with regions where the superconducting transitions is narrowing.

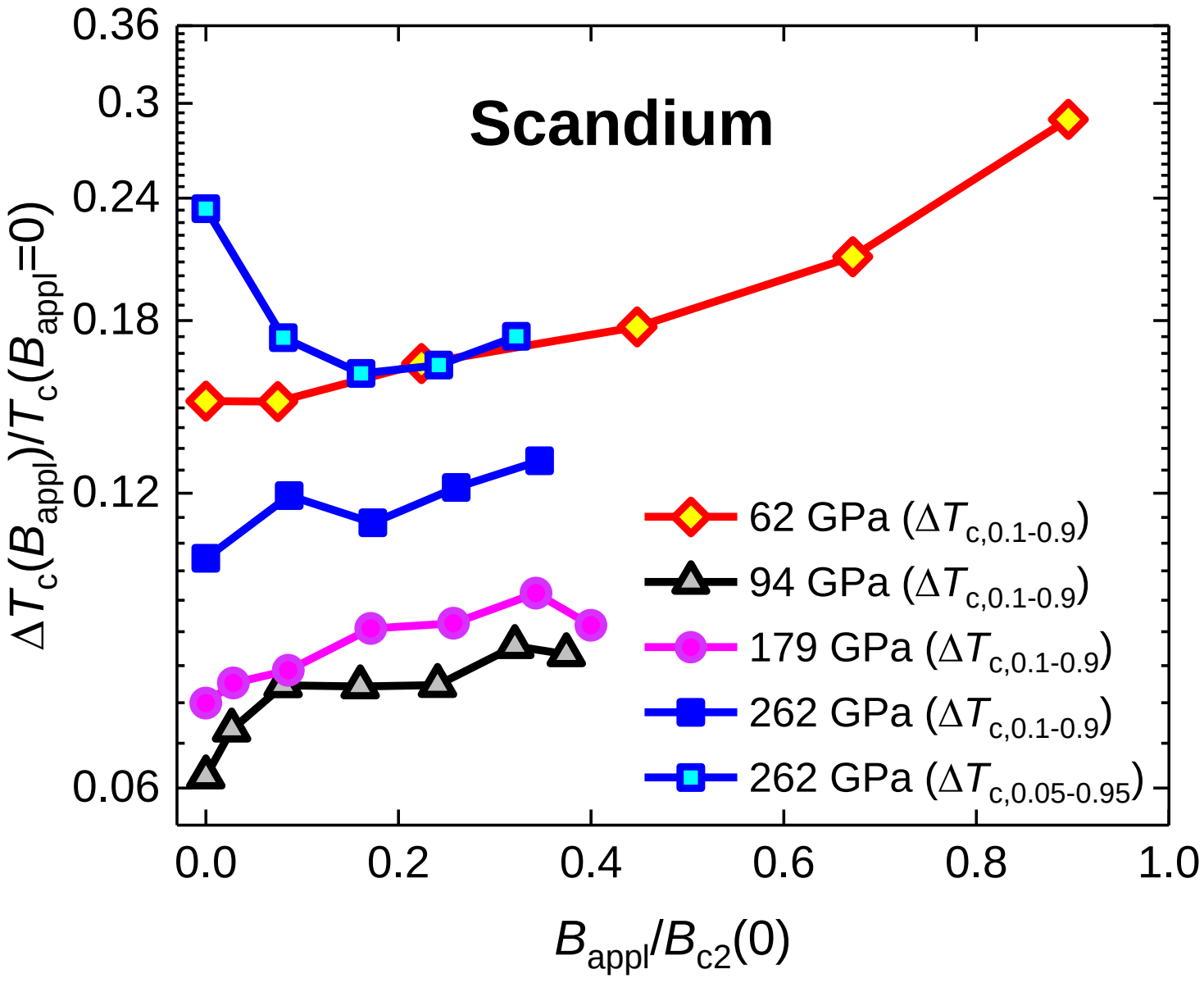

**Figure 4.** Transition width, $\frac{\Delta T_c(B_{appl})}{T_c(B_{appl}=0)}$, *vs.* applied magnetic field, $\frac{B_{appl}}{B_{c2}(0)}$, for highly compressed elemental scandium.

## VI. Hydride superconductors.

In this section, we focus on hydrogen-rich compounds at high pressures. In Figure 5, we present several $\frac{\Delta T_c(B_{appl})}{T_c(B_{appl}=0)}$ *vs.* $\frac{B_{appl}}{B_{c2}(0)}$ relationships for these superconductors. Notably, these relationships exhibit contrasting trends compared to the claims made in Ref.[1], where it was suggested that in all highly compressed hydrides, the "*broadening either stays the same as for H=0 or even decreases with applied field*".

To illustrate the similarity between the $\frac{\Delta T_c(B_{appl})}{T_c(B_{appl}=0)}$ *vs.* $\frac{B_{appl}}{B_{c2}(0)}$ relationships for highly compressed hydrides and materials for which the superconducting state remains undisputed by the authors[1], we included data for $MgB_2$+SiC sample (sample C from Ref.[5]) and $CeH_9$ sample at 137 GPa from Ref.[11] in Figure 5. It is evident that for $\frac{B_{appl}}{B_{c2}(0)} \lesssim 0.07$, the transitions exhibit a narrowing trend, whereas at higher applied magnetic fields, the superconducting transition width increases.

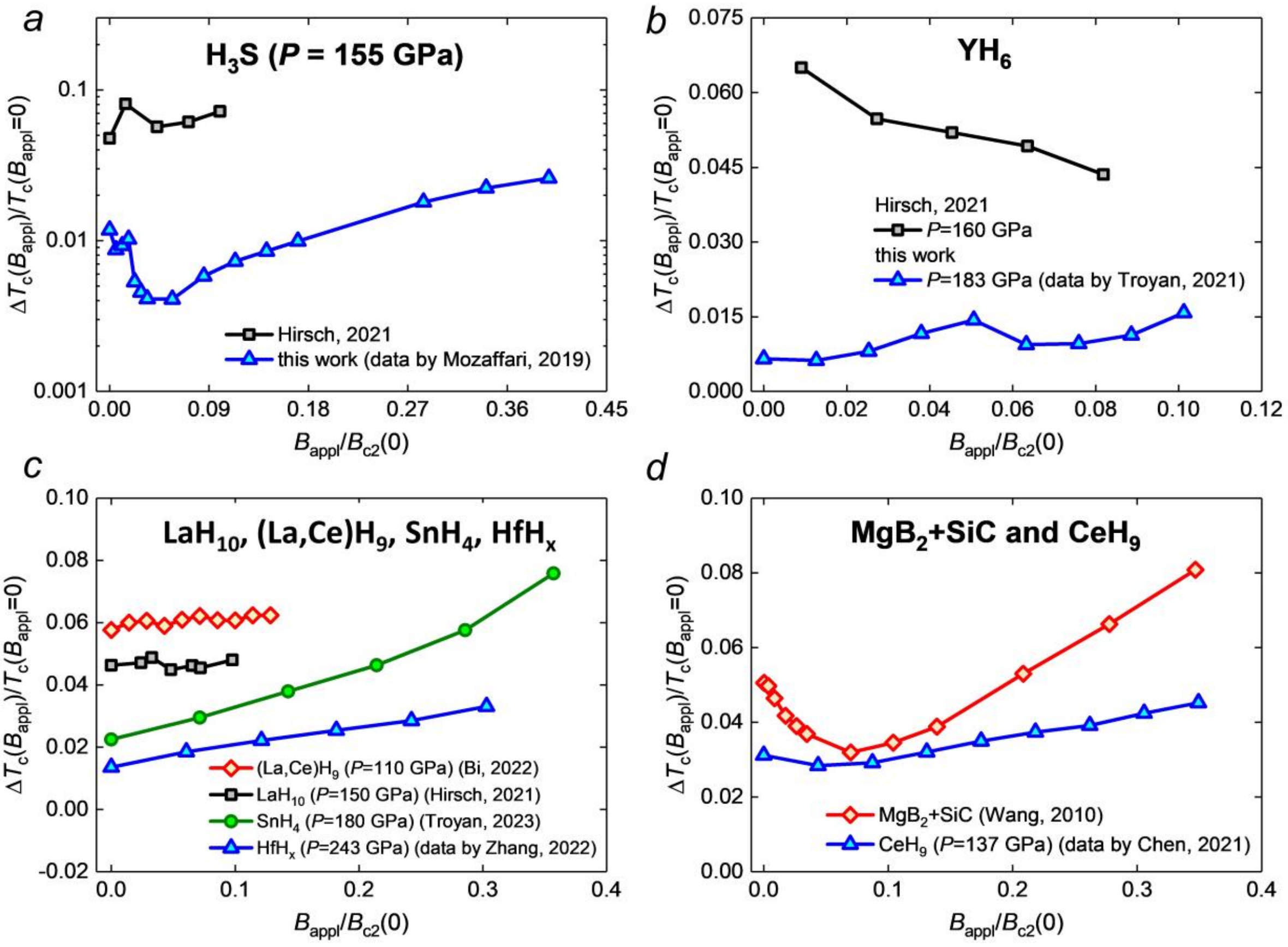

**Figure 5.** Transition width, $\frac{\Delta T_c(B_{appl})}{T_c(B_{appl}=0)}$, *vs.* applied magnetic field, $\frac{B_{appl}}{B_{c2}(0)}$, for highly-compressed hydrides. Dependences reported in Ref.[1] are in black. (a) $H_3S$ at 155 GPa: blue triangles indicate data points derived from $R(T, B_{appl}, P = 155\ GPa)$ data reported by Mozaffari *et al*.[12] For additional details see *Appendix E* in Supplementary Materials. (b) $YH_6$ at 166 GPa: the data (blue triangles) are deduced from $R(T, B_{appl}, P = 166\ GPa)$ reported by Troyan *et al.[31]* Additional details can be found in *Appendix F* in Supplementary Materials. (c) $LaH_{10}$ at 150 GPa (black squares from Ref.[1]); $(La,Ce)H_9$ at 110 GPa (red diamonds from Ref.[32], details in *Appendix G* in Supplementary Materials); $SnH_4$ at 180 GPa (green circles from Ref.[33], details in *Appendix H* in Supplementary Materials); $HfH_x$ at 243 GPa (blue triangles from $R(T, B_{appl}, P = 166\ GPa)$ reported by Zhang *et al.[34]*, details in *Appendix I* in Supplementary Materials. (d) $MgB_2$ doped by SiC: red diamonds correspond to sample C from Ref.[5] Also included is $CeH_9$ at 137 GPa (blue triangles from Ref.[11], details in *Appendix J* in Supplementary Materials).

## VII. Magnetic phase diagram for superconductors.

In their work[1], the authors also provided an extended narrative on magnetic properties of superconductors. In their Figure 1, they presented "*Schematic phase diagrams of type-I and type-II superconductors in a magnetic field. N = normal state, S = superconducting state*". We have redrawn their Figure 1a from Ref.[1], which pertains to type-I superconductors, and included it in Figure 6a.

We argue that one of the fundamental states of type-I superconductors, the intermediate state, was overlooked by the authors in their Figure 1a. Additionally, this state was omitted from the entire text in Ref.[1].

The intermediate state of a type-I superconductor is characterized by the coexistence of the superconducting and the normal states within the sample. A thorough description of this fundamental state of type-I superconductors, along with the detailed review of experimental data, can be found in the book by Poole *et al*.[35] (refer to Chapter 11 therein).

In Figure 6b and 6c, we have schematically depicted phase diagrams for sphere-shaped (panel b) and disk-shaped (panel c) samples of a type-I superconductor.

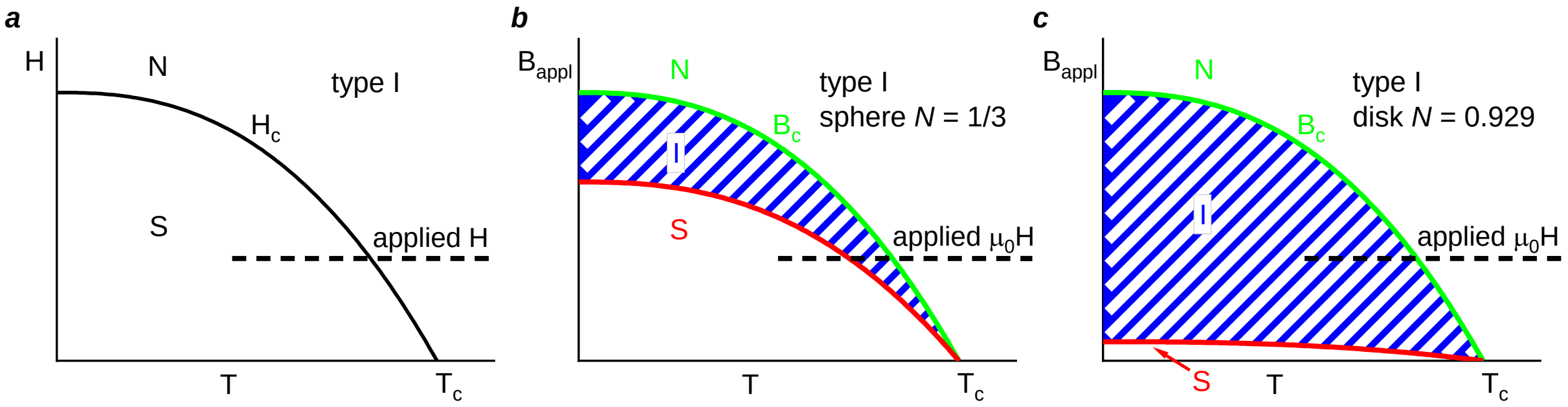

**Figure 6.** Schematic phase diagrams of a type-I superconductor. N represents normal state, S - superconducting state, I - intermediate state, and *N* - the demagnetization factor. (a) The diagram published by the authors in Ref.[1] (b) The diagram for a spherical sample of a type-I superconductor. (c) The diagram for a disk-shaped sample of a type-I superconductor with sample dimensions typical for hydride samples in diamond anvil cell.

As all near-room temperature hydride samples have a thin-disk shape, the phase diagram in Figure 6c was simulated based on the corresponding demagnetization factor, *N*, calculated using a routine proposed by Brandt[36]:

$$B_c^{disk} = B_c \times (1 - N) = B_c \times \left(1 - \left(1 - \frac{1}{1 + q \times \frac{a}{b}}\right)\right) \tag{2}$$

$$q = \frac{4}{3\pi} + \frac{2}{3\pi} tanh\left(1.27 \times \frac{b}{a} \times ln\left(1 + \frac{a}{b}\right)\right) \tag{3}$$

where 2$a$ is the disk diameter and 2$b$ is the disk thickness. For the calculations in Figure 6c, we used typical values for hydride samples in a diamond anvil cell[37]: 2$a$ = 80 μm, 2$b$ = 2.8 μm.

Given that the primary magnetic state of type-I superconductors was overlooked in Ref.[1], we advocate for further examination of recent papers published by the same authors on magnetic properties of highly compressed hydrides and other type-II superconductors[16-18, 38-42].

## VIII. Equation for $R(T,B_{appl})$ transition curve.

We also need to mention that Eq. 26[1] has several unavoidable problems. Details of these problems have been described in Ref.[6].

## Conclusion

In this Comment, we argue that the primary conclusions drawn by the authors[1], specifically their assertion that "*if hydrides under pressure are truly superconductors, they do not obey the same physical principles that all other type-II superconductors, whether conventional or unconventional, obey.*" are based on the following flawed approaches:

(1) The utilization of undescribed and unknown method(s) for data processing and analysis;
(2) Data selectivity bias, where the chosen samples for analysis do not adequately represent the general cases for both non-hydrides and hydride superconductors;
(3) An incorrect representation of the primary magnetic properties of superconductors.

Additionally, we emphasized that our analysis supports the view that the $\frac{\Delta T_c(B_{appl})}{T_c(B_{appl}=0)}$ *vs.* $\frac{B_{appl}}{B_{c2}(0)}$ trend cannot be used as the criterion for determining the existence or absence of superconductivity in any material, as detailed in Ref.[6].

**Acknowledgement**

EFT thanks the financial support provided by the Ministry of Science and Higher Education of Russia (theme “Pressure” No. 122021000032-5).

## Supplementary Materials to
## Broadening of In-Field Superconducting Transitions in Hydrides.
## Comment on "Nonstandard Superconductivity or No Superconductivity in Hydrides under High Pressure" [J. E. Hirsch and F. Marsiglio, Phys. Rev. B **103**, 134505 (2021)]

E. F. Talantsev[1], V. S. Minkov[2], F. F. Balakirev[3], M. I. Eremets[2]

[1]*M. N. Mikheev Institute of Metal Physics, S. Kovalevskoy St. 18, Ekaterinburg 620108, Russian Federation*
[2]*Max Planck Institute for Chemistry; Hahn-Meitner-Weg 1, Mainz 55128, Germany*
[3]*NHMFL, Los Alamos National Laboratory, MS E536, Los Alamos, New Mexico 87545, USA*

### Appendix A. $MgB_2$

In their Figure 5, Acharya *et al.*[1] presented $R(T, B_{appl,perp})$ and $R(T, B_{appl,para})$ curves, where the designations $B_{appl,perp}$ and $B_{appl,para}$ indicate the applied field directed perpendicular and parallel to the (0001) plane of the $MgB_2$ film, respectively. Figure A1 displays these $R(T, B_{appl,perp})$ and $R(T, B_{appl,para})$ curves, along with the corresponding data points for deduced $T_{c,0.9}(B_{appl})$ and $T_{c,0.1}(B_{appl})$.

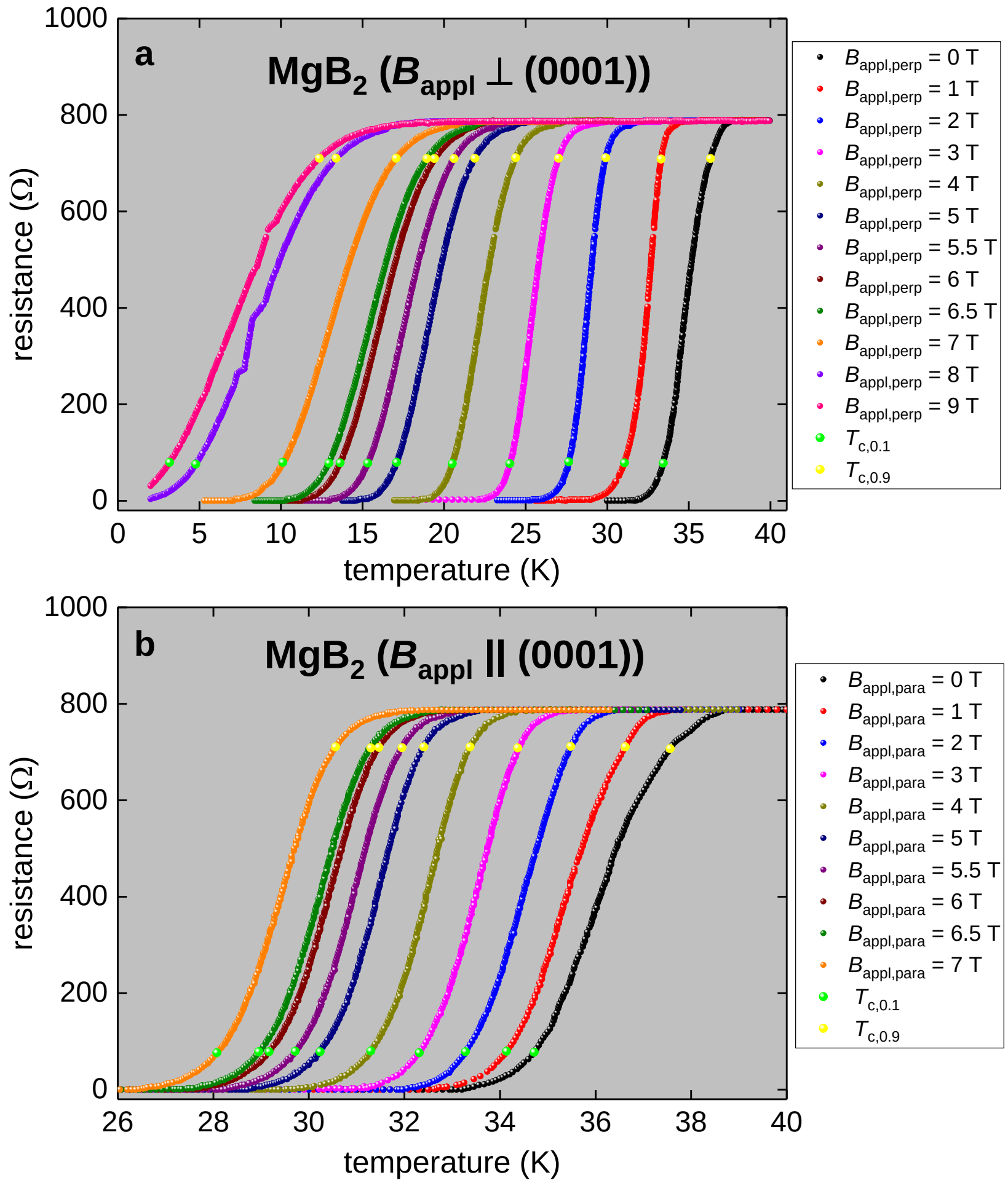

**Figure A1.** Resistive curves $R(T, B_{appl})$ and deduced $T_{c,0.9}(B_{appl})$ (yellow circles) and $T_{c,0.1}(B_{appl})$ (green circles) for an $MgB_2$ thin film in an applied magnetic field directed (a) perpendicular and (b) parallel to the (0001) plane of the epitaxial film. The raw $R(T, B_{appl})$ curves were originally reported by Acharya *et al.*[1] in their Figure 5.

In Figure A2, we presented the $\frac{\Delta T_c(B_{appl,perp})}{T_c(B_{appl}=0)}$ *vs.* $\frac{B_{appl,perp}}{B_{c2,perp}(0)}$ and $\frac{\Delta T_c(B_{appl,para})}{T_c(B_{appl}=0)}$ *vs.* $\frac{B_{appl,para}}{B_{c2,para}(0)}$ relationships for the $MgB_2$ film, along with the data reported by Hirsch and Marsiglio[2] for $MgB_2$ (the authors extracted their dataset from experimental data reported by Canfield *et al.*[3]). To calculate the values in Figure A2, we used $B_{c2,perp}(0) = 10\ T$ (which can be obtained from $B_{c2,perp}(T)$ data reported by Acharya *et al.*[1] in their Figure 5(c)) and $B_{c2,para}(0) = 26\ T$, which can be calculated using the upper critical field anisotropy, $\gamma_{B_{c2}(0)} = \frac{B_{c2,para}(0)}{B_{c2,perp}(0)} = 2.6$, as reported by Xu *et al.*[4]. It´s worth noting that Lee *et al.*[5] reported a very similar value of $\gamma_{B_{c2}(0)} = 2.7$.

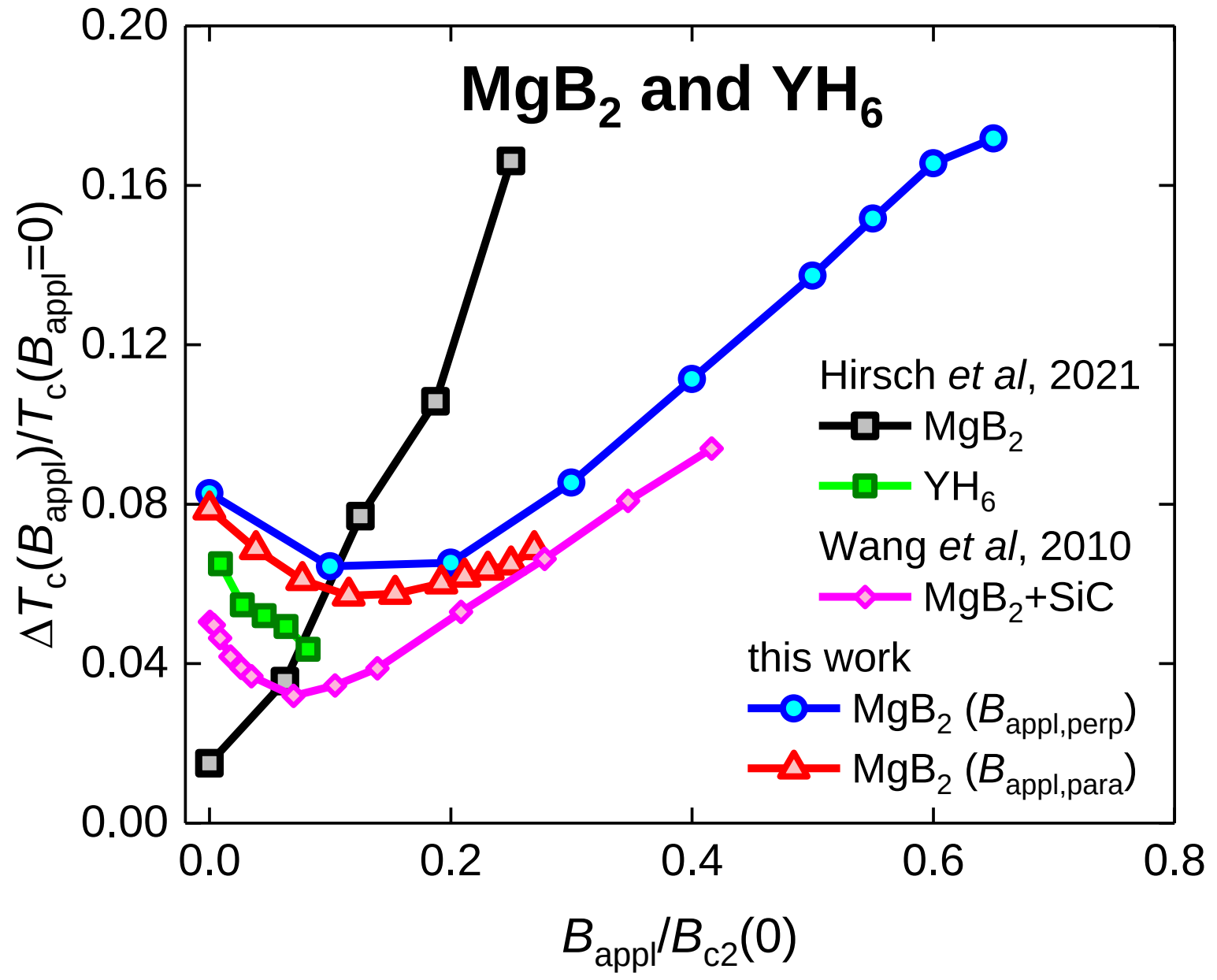

**Figure A2.** $\frac{\Delta T_c(B_{appl})}{T_c(B_{appl}=0)}$ *vs.* $\frac{B_{appl}}{B_{c2}(0)}$ relationships for $MgB_2$ and highly-compressed $YH_6$. Data for $MgB_2$: black squares (originally reported by Hirsch and Marsiglio[2], deduced from data in Ref.[3]); blue circles and red triangles are data deduced in this work from the data reported in Ref.[1] (refer to Figure A1(a) and A1(b) for details); magenta diamonds are data recalculated in this work from the data reported for sample C ($MgB_2$+SiC) in Ref.[6]. Green squares represent data for highly-compressed $YH_6$ (as reported by Hirsch and Marsiglio[2], deduced from Ref.[7]).

One can observe from Figure A2 that the dependence of $\frac{\Delta T_c(B_{appl})}{T_c(B_{appl}=0)}$ *vs.* $\frac{B_{appl}}{B_{c2}(0)}$ claimed by Hirsch and Marsiglio[2] for $MgB_2$ differs fundamentally from the dependences that we deduced from experimental $R(T, B_{appl})$ data reported elsewhere[1]. For instance, at low applied magnetic fields, $\frac{B_{appl}}{B_{c2}(0)} \lesssim 0.15$, we observe a narrowing of the transition width with increasing applied field (see Figure A2). The effect of transition width broadening *vs.* applied field is observed for $\frac{B_{appl}}{B_{c2}(0)} \gtrsim 0.15$, however the rate of the broadening is three-fold lower than the one reported by Hirsch and Marsiglio[2].

In Figure A2, we also presented the $\frac{\Delta T_c(B_{appl})}{T_c(B_{appl}=0)}$ *vs.* $\frac{B_{appl}}{B_{c2}(0)}$ relationship for highly compressed yttrium superhydride, $YH_6$, for which raw data was reported in Ref.[7] It´s worth noting that the authors[2] used the wrong designation for this phase, i.e., $YH_9$[2] instead of the correct $YH_6$[7], in their Figures 7 and 11[2]. Additionally, they incorrectly reported the sample pressure (the pressure was 160 GPa[7], whereas in Figure 11 in Ref.[2], the pressure is indicated as 185 GPa).

To address the possible argument that in Figure 3a we presented data for the $MgB_2$ epitaxial highly oriented thin film, while the $MgB_2$ sample analyzed by Hirsch and Marsiglio[2] is a polycrystalline sample, in Figure 3 we also displayed the dependence reported by Wang *et al.*[6] for polycrystalline $MgB_2$ samples doped by SiC, for which $\Delta T_c(B_{appl})$ was defined by Equation 1.

The effect of the $\Delta T_c(B_{appl})$ narrowing at low applied magnetic fields, $B_{appl}$, was reported by Wang *et al.*[6] for undoped $MgB_2$ (see Figure 1 in Ref.[6]), and for doped $MgB_2$ samples (see Figures 2 and 3 in Ref.[6]). In our Figures 3b and A2, we recalculated $\Delta T_c(B_{appl})$ vs. $B_{appl}$ data (as shown by Wang *et al.* in their Figure 3[6]) in the form of $\frac{\Delta T_c(B_{appl})}{T_c(B_{appl}=0)}$ *vs.* $\frac{B_{appl}}{B_{c2}(0)}$ using $T_c(B_{appl}=0) = 39.2\ K$ and $B_{c2}(0) = 28.9\ T$, as reported in Ref.[6]

We find that $YH_6$, $MgB_2$, and $MgB_2$+SiC exhibit nearly identical rates of narrowing in the $\frac{\Delta T_c(B_{appl})}{T_c(B_{appl}=0)}$ *vs.* $\frac{B_{appl}}{B_{c2}(0)}$ dependence within the field range of $\frac{B_{appl}}{B_{c2}(0)} \lesssim 0.08$ (Figure A2).

Based on this observation, the trend of transition width narrowing *vs.* applied field in $YH_6$, which was claimed in Ref.[2] as experimental evidence for "*nonstandard superconductivity or no superconductivity*" in highly-compressed hydrides, is actually observed in the "*standard*" $MgB_2$ superconductor (see Figures 3a and 3b, and A2).

Thus, we demonstrated that the available experimental data for $MgB_2$ contradicts the primary claim of the authors[2]:

"*For all standard superconductors… the transition broadens under application of a magnetic field, as expected from the theory discussed in the next section. We looked at many more examples of standard superconductors and the qualitative behavior is always the same. Note that the upward slope is larger for the higher Tc cases, YBCO (unconventional) and $MgB_2$ (conventional), as expected from the standard theory discussed in the next section…*

*Our finding is remarkable because hydride superconductors are considered to be textbook examples of conventional BCS-Eliashberg superconductors driven by the electron-phonon interaction, which unlike unconventional superconductivity, has been thought to be well understood for the last 60 years. All the theoretical analysis of these materials conclude that their $T_c$ is perfectly well explained by standard superconductivity theory... How is it possible then that the behavior of resistivity in the presence of a magnetic field does not follow the standard behavior seen in type-II superconductors* [8]*?*"

In contrast, we argue that the specific $\frac{\Delta T_c(B_{appl})}{T_c(B_{appl}=0)}$ *vs.* $\frac{B_{appl}}{B_{c2}(0)}$ relationships presented by Hirsch and Marsiglio[2] for $MgB_2$ do not represent a general case for this superconductor, as there is no universal broadening $\frac{\Delta T_c(B_{appl})}{T_c(B_{appl}=0)}$ *vs.* $\frac{B_{appl}}{B_{c2}(0)}$ trend for this type-II conventional superconductor, as the authors claimed.

The available experimental data presented for $MgB_2$ herein (see Figures A1 and A2) provide us with grounds to propose that *this dependence is determined not by sample composition, but by the sample manufacturing technique, which creates a unique sample microstructure. This microstructure, in turn, determines the distributions of local $T_c$ and $B_{c2}$ at the nanoscale level within the entire sample. These unique $T_c$ and $B_{c2}$ distributions, in turn, determine the particular shape, sharpness, and in-field dependence of the transition width.*

**Appendix B. Pnictide superconductor $Na(Fe_{0.99}Co_{0.01})As$**

Hirsch and Marsiglio[2] also presented the $\frac{\Delta T_c(B_{appl,perp})}{T_c(B_{appl}=0)}$ *vs.* $\frac{B_{appl,perp}}{B_{c2,perp}(0)}$ relationships for the $Ba(Fe_{0.645}Ru_{0.355})_2As_2$ single crystal, for which the raw $R(T, B_{appl,perp})$ data were reported by Sharma *et al.*[9]. The authors[2] claimed that this iron-based superconductor is "*standard*" and exhibits the $\frac{\Delta T_c(B_{appl,perp})}{T_c(B_{appl}=0)}$ broadening trend *vs.* $\frac{B_{appl,perp}}{B_{c2,perp}(0)}$, which is the "*empirical rule*" observed in "*all standard type-II superconductors*" (the subscript *perp* designates the direction of the applied magnetic field in the perpendicular direction to the (001) crystallographic plane of the single crystal). The authors did not provide the reasons why this compound was chosen[2] to be a representative compound for the entire iron-based superconductors family

In Figure A3, we plot $R(T, B_{appl,perp})$ data reported for the Na($Fe_{0.99}Co_{0.01}$)As iron-based superconductor by Choi *et al.*[10], and we highlight the $T_{c,onset}(B_{appl,perp})$, $T_{c,0.9}(B_{appl,perp})$, and $T_{c,0.1}(B_{appl,perp})$ data points in each curve.

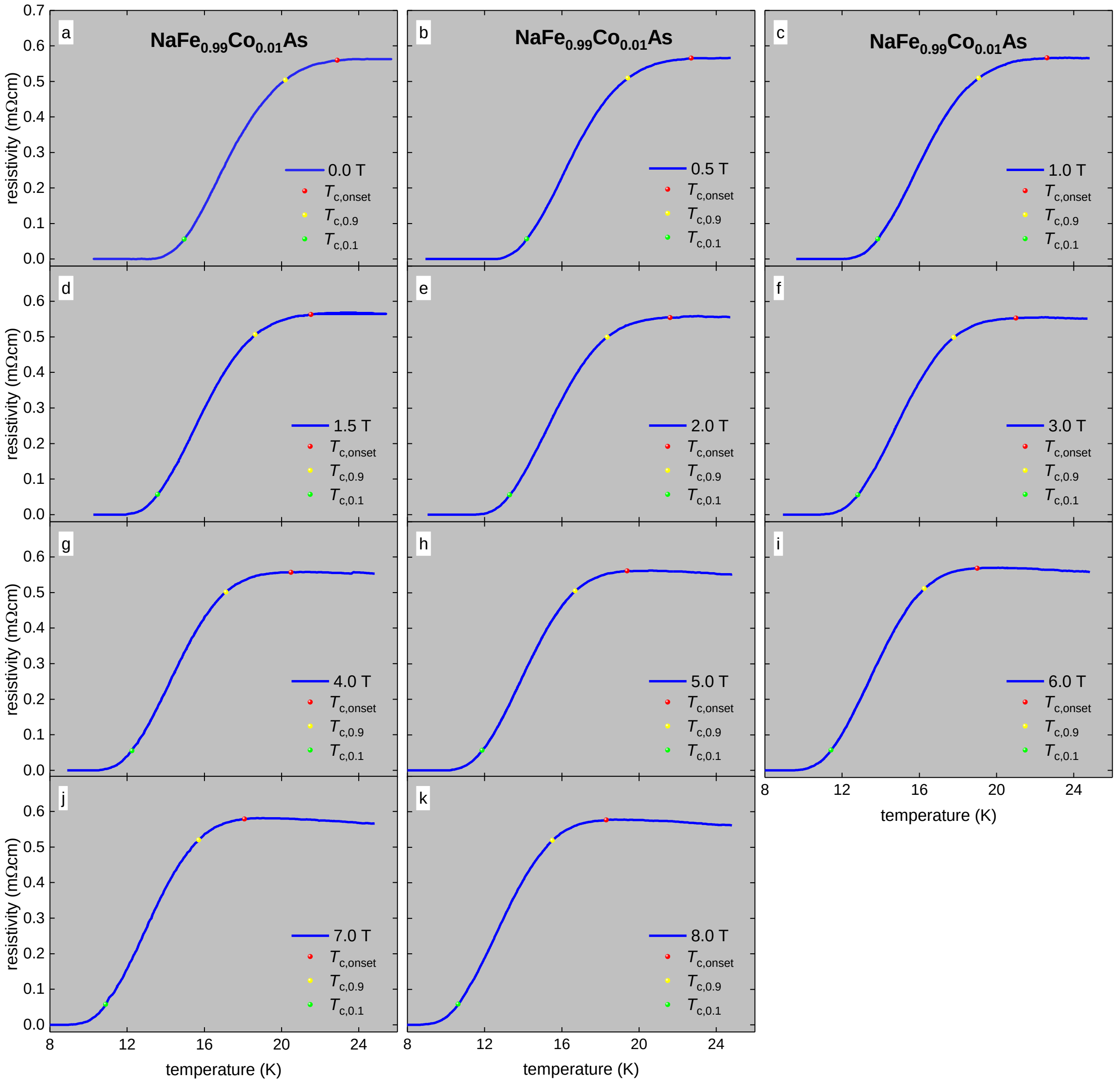

**Figure A3.** Resistivity curves $\rho(T, B_{appl,perp})$ and deduced $T_{c,onset}(B_{appl,perp})$ (red circles), $T_{c,0.9}(B_{appl,perp})$ (yellow circles) and $T_{c,0.1}(B_{appl,perp})$ (green circles) for the Na($Fe_{0.99}Co_{0.01}$)As single crystal. Raw $\rho(T, B_{appl,perp})$ data were reported by Choi *et al.*[10] in their Figure 4(a).

These datasets were used to calculate the $\frac{\Delta T_c(B_{appl,perp})}{T_c(B_{appl}=0)}$ *vs.* $\frac{B_{appl,perp}}{B_{c2,perp}(0)}$ relationships for the $Na(Fe_{0.99}Co_{0.01})As$ iron-based superconductor using Equation 1 in Figure A4, where we used $B_{c2,perp}(0) = 25.5\ T$ as reported for this material in the original work[10].

In Figure A4, we also presented the $\frac{\Delta T_c(B_{appl})}{T_c(B_{appl}=0)}$ *vs.* $\frac{B_{appl}}{B_{c2}(0)}$ relationships reported in Ref.[2] for $Ba(Fe_{0.645}Ru_{0.355})_2As_2$ and $(La,Y)H_{10}$.

It´s worth noting that $Na(Fe_{0.99}Co_{0.01})As$ does not conform to the behaviour claimed to be universal for all "*standard superconductors*", i.e., "*the broadening either stays the same as for H=0 or even decreases with applied field*"[2].

Thus, based on the criterion proposed by Hirsch and Marsiglio[2], the $Na(Fe_{0.99}Co_{0.01})As$ single crystal could be classified as exhibiting "*nonstandard superconductivity or no superconductivity*".

In addition, a comparison of the $\frac{\Delta T_c(B_{appl})}{T_c(B_{appl}=0)}$ *vs.* $\frac{B_{appl}}{B_{c2}(0)}$ trends for $Na(Fe_{0.99}Co_{0.01})As$ and $(La,Y)H_{10}$ (the latter trend is taken from Ref.[2]) shown in Figure A4 leads to the conclusion that $(La,Y)H_{10}$ is more likely to be classified as a "*standard superconductor*" because the $\frac{\Delta T_c(B_{appl})}{T_c(B_{appl}=0)}$ slightly increases with an applied magnetic field within the analyzed field range.

Another important issue associated with Ref.[2] is that the authors compared the ambient pressure superconductors on one hand with highly-compressed hydrides on the other hand. *De facto*, this comparison excludes all highly-compressed non-hydride superconductors from consideration. We address this gap in the following sections.

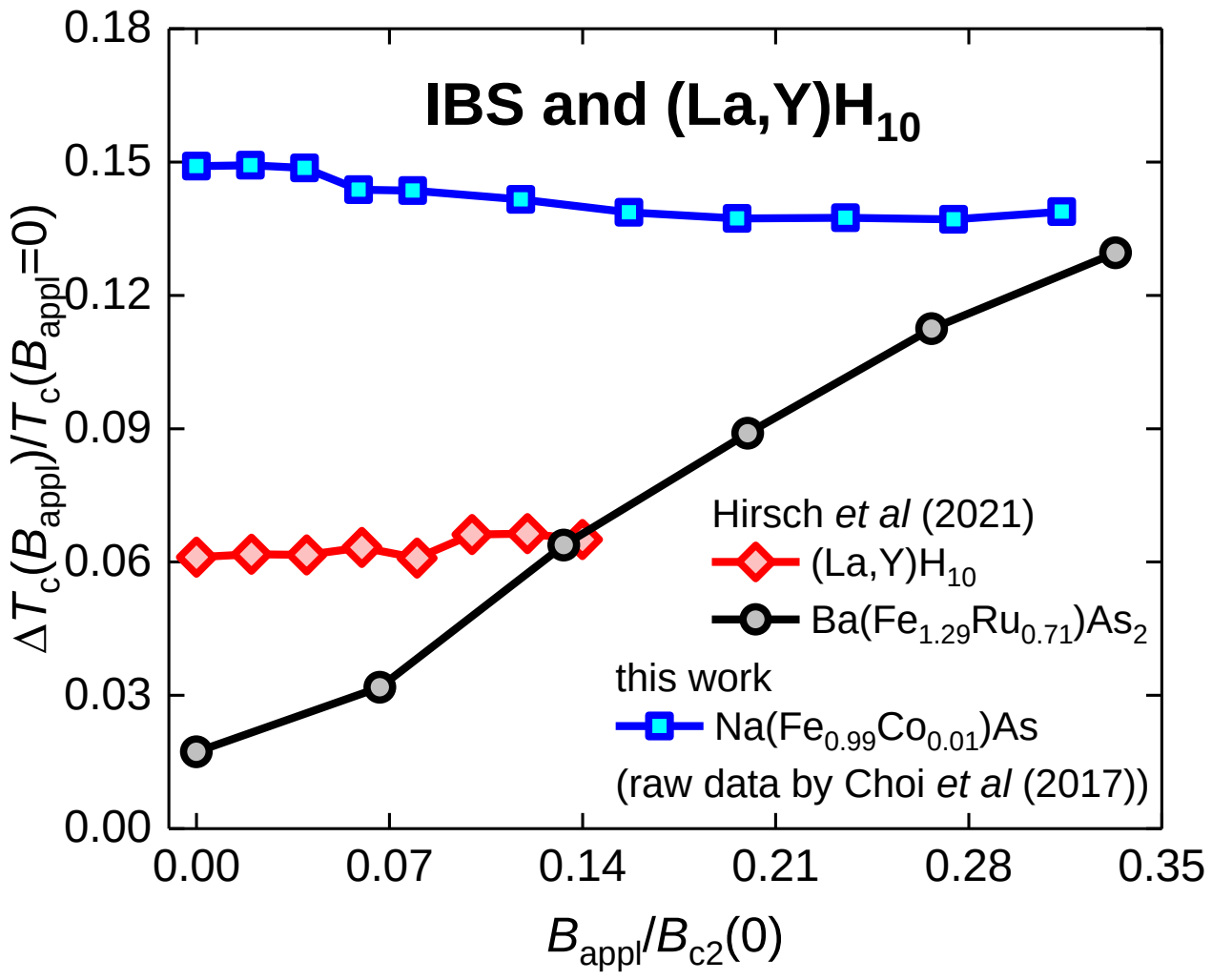

**Figure A4.** $\frac{\Delta T_c(B_{appl})}{T_c(B_{appl}=0)}$ *vs.* $\frac{B_{appl}}{B_{c2}(0)}$ for the $Na(Fe_{0.99}Co_{0.01})As$ single crystal (the raw data were reported by Choi *et al.*[18]) (blue squares), $Ba(Fe_{0.645}Ru_{0.355})_2As_2$ single crystal (reported in Ref.[2] to be deduced from data by Sharma *et al.*[9] (black circles), and $(La,Y)H_{10}$ (reported in Ref.[2] to be deduced from data reported by Semenok *et al.*[11]) (red diamonds).

### Appendix C. Highly-compressed $Bi_2Sr_2CaCu_2O_8$

In Figure A5, we presented the evolution of $R_{ab}(T, B_{appl}, P)$ in the high-temperature superconducting cuprate $Bi_2Sr_2CaCu_2O_8$ subjected to high pressure. The raw $R_{ab}(T, B_{appl}, P)$ data was reported by Guo *et al.*[12]

In Figure A6, we show the $R_{ab}(T, B_{appl,perp}, P = 9.0\ GPa)$ data for $Bi_2Sr_2CaCu_2O_8$, where we highlighted the $T_{c,onset}(B_{appl,perp}, P = 9.0\ GPa)$, $T_{c,0.9}(B_{appl,perp}, P = 9.0\ GPa)$, and $T_{c,0.1}\left(B_{appl,perp}, P = 9.0\ GPa\right)$ data points in each curve. These datasets were used to calculate the $\frac{\Delta T_c(B_{appl,perp})}{T_c(B_{appl}=0)}$ *vs.* $\frac{B_{appl,perp}}{B_{c2,perp}(0)}$ relationships for $Bi_2Sr_2CaCu_2O_8$ cuprate using Equation 1 (see Figure A7).

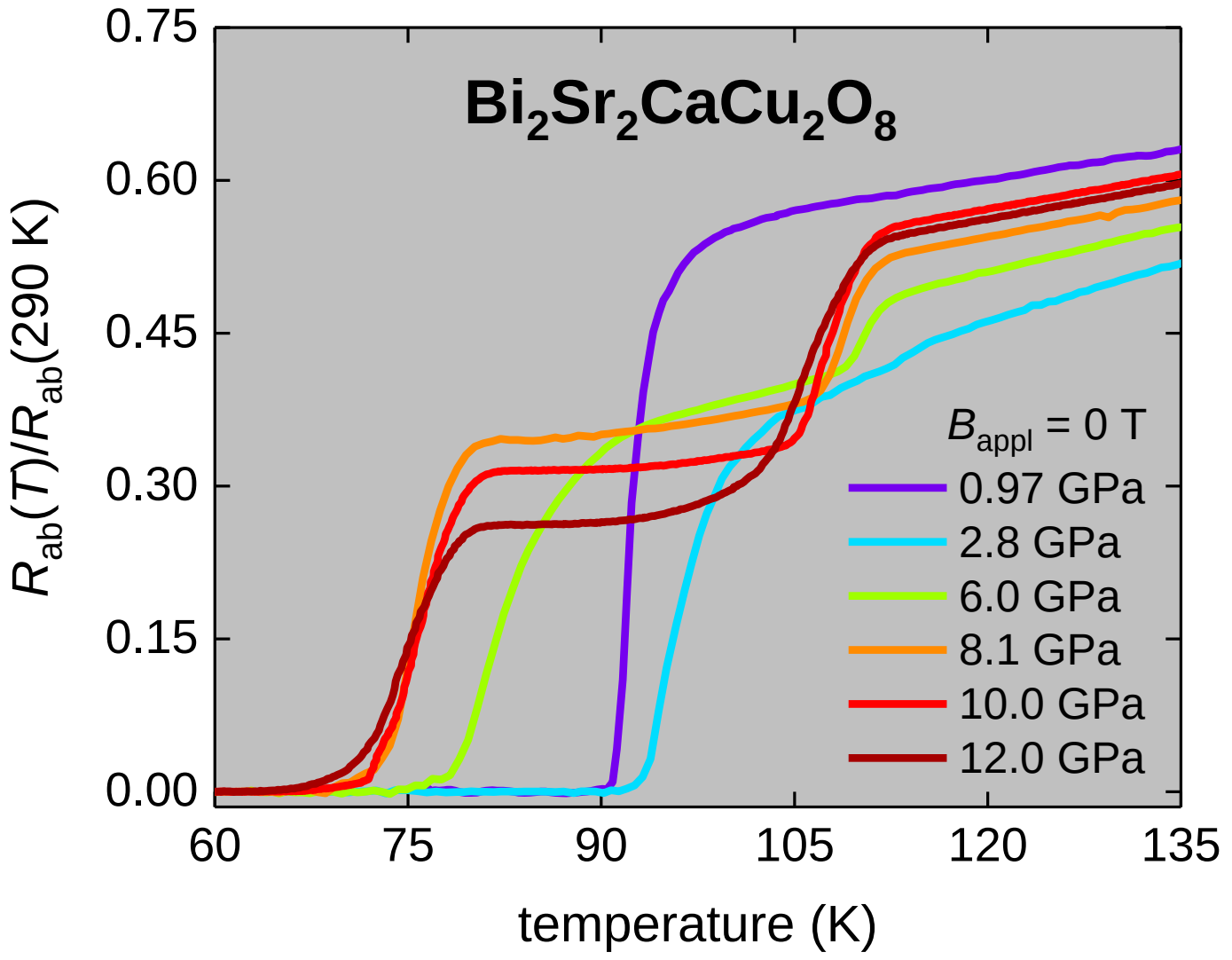

**Figure A5.** Evolution of $\frac{R_{ab}(T, B_{appl}=0\ T, P)}{R_{ab}(T=290\ K, B_{appl}=0\ T, P)}$ for the single crystal $Bi_2Sr_2CaCu_2O_8$ (the raw data were reported by Guo *et al.*[12]).

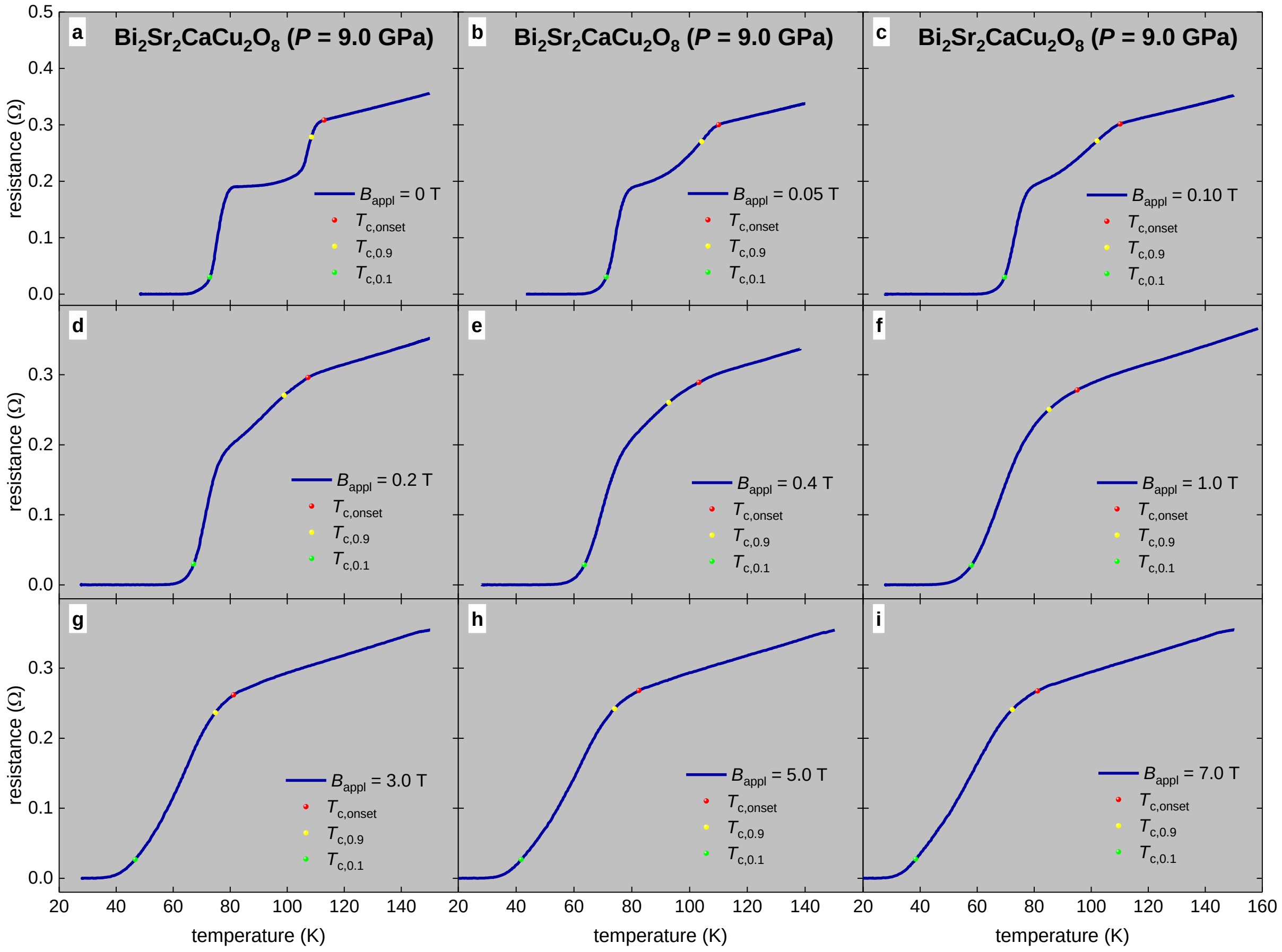

**Figure A6.** Resistance curves $R\left(T, B_{appl,perp}\right)$ and deduced $T_{c,onset}\left(B_{appl,perp}\right)$ (red circles), $T_{c,0.9}\left(B_{appl,perp}\right)$ (yellow circles), and $T_{c,0.1}\left(B_{appl,perp}\right)$ (green circles) for the $Bi_2Sr_2CaCu_2O_8$ single crystal compressed at 9 GPa. The raw $R\left(T, B_{appl,perp}\right)$ data were reported by Guo *et al.*[12]

It should be stressed that Hirsch and Marsiglio[2] refuted the existence of superconductivity in $YH_6$ based on a significantly less pronounced double-step transition curve $R(T, B_{appl}, P = 160\ GPa)$ [7] when compared to the transition in Ref. [12]. Following this logic, these authors should also refute the existence of superconductivity in highly-compressed $Bi_2Sr_2CaCu_2O_8$ because the transition width, $\frac{\Delta T_c(B_{appl})}{T_c(B_{appl}=0)}$, is also a decreased function of the applied magnetic field in the field range of $\frac{B_{appl}}{B_{c2}(0)} < 0.08$ (see Figure A7).

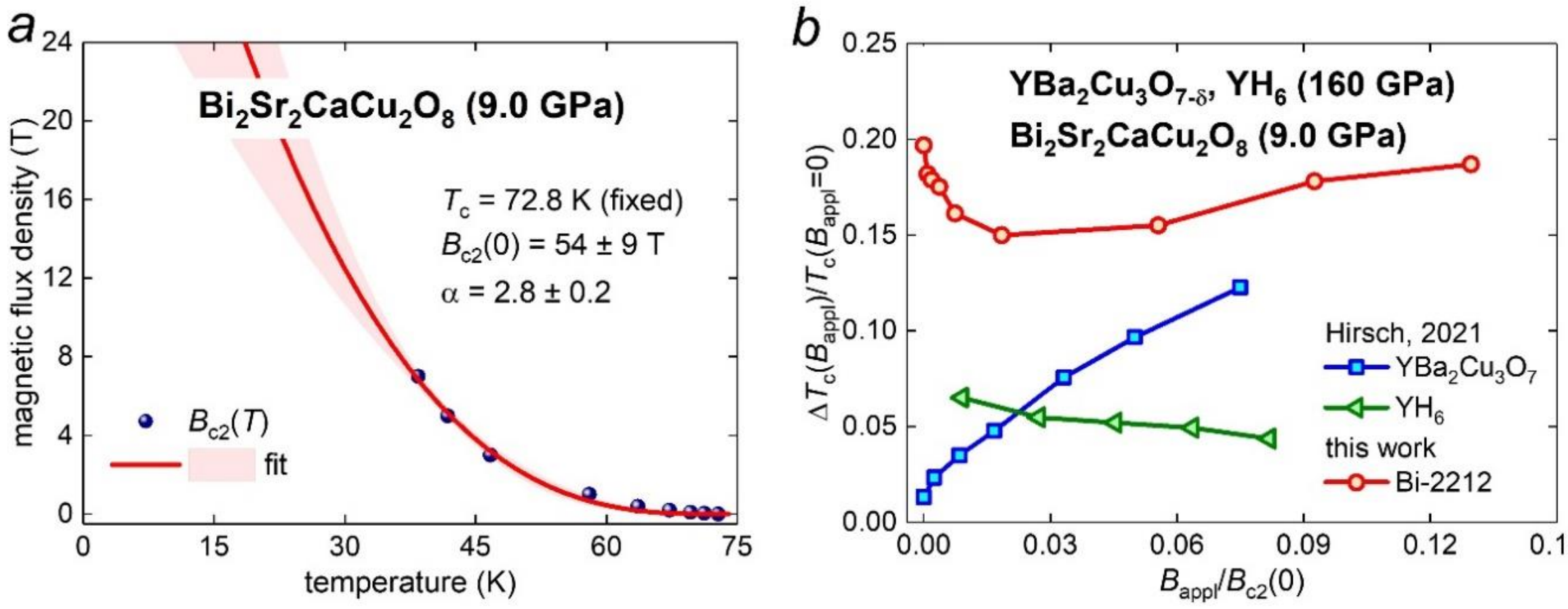

**Figure A7.** (a) The upper critical field, $B_{c2}(T)$, deduced by the criterion of $T_{c,0.1}(B_{appl}, P) = \frac{R_{ab}(T,B_{appl},P)}{R_{ab}(T_c^{onset},B_{appl},P)} = 0.10$, and data fit to Equation E2 for $Bi_2Sr_2CaCu_2O_8$ compressed at 9 GPa. The raw $R_{ab}(T, B_{appl}, P)$ data were reported by Guo *et al.*[12]. (b) Dependences of the transition width, $\frac{\Delta T_{c,0.9-0.1}(B_{appl},P=9.0\ GPa)}{T_c(B_{appl}=0,P=9.0\ GPa)}$, *vs.* applied magnetic fields, $\frac{B_{appl}}{B_{c2}(0,P=9.0\ GPa)}$, for highly compressed $Bi_2Sr_2CaCu_2O_8$ (red circles, this work), and the $\frac{\Delta T_c(B_{appl})}{T_c(B_{appl}=0)}$ *vs.* $\frac{B_{appl}}{B_{c2}(0)}$ dependence for $YBa_2Cu_3O_{7-\delta}$ (blue squares, from Ref.[2]) and $YH_6$ (green triangles, from Ref.[2]). 95% confidence bands are shown by pink shadow areas.

To plot $\frac{\Delta T_{c,0.9-0.1}(B_{appl},P=9.0\ GPa)}{T_c(B_{appl}=0,P=9.0\ GPa)}$ *vs.* $\frac{B_{appl}}{B_{c2}(0,P=9.0\ GPa)}$ for the $Bi_2Sr_2CaCu_2O_8$ single crystal, we fitted $B_{c2}(T, P = 9.0\ GPa)$ values (defined by the $\frac{R(T,B_{appl},P)}{R(T_c^{onset},B_{appl},P)} = 0.10$ criterion) to an analytical extrapolative expression similar to one proposed by Grissonnanche *et al.*[13]:

$$B_{c2}(T) = B_{c2}(0)\left(1 - \frac{T}{T_c}\right)^{\alpha} \quad \text{(S1)}$$

where $\alpha$ is a free fitting parameter. The deduced values are $B_{c2}(0) = 54 \pm 9\ T$ and $\alpha = 2.8 \pm 0.2$ (see Figure A7(a)).

In Figure A7(b), we also presented the $\frac{\Delta T_c(B_{appl})}{T_c(B_{appl}=0)}$ *vs.* $\frac{B_{appl}}{B_{c2}(0)}$ relationship for $YBa_2Cu_3O_{7-\delta}$ superconductor, which was reported by Hirsch and Marsiglio[2] as the representative example for cuprate superconductors. However, it is evident from Figure A7 that highly-compressed $Bi_2Sr_2CaCu_2O_8$ exhibits a substantially different $\frac{\Delta T_c(B_{appl})}{T_c(B_{appl}=0)}$ *vs.* $\frac{B_{appl}}{B_{c2}(0)}$ dependence.

**Appendix D. Highly-compressed elemental scandium**

From a variety of reports on superconductivity in highly pressurized pure metals, we chosen recent study by Ying *et al.*[14], who reported the record high onset of the superconducting transition for elemental scandium, with $T_{c,onset}(B_{appl} = 0) = 36\ K$ at pressures of 215 – 262 GPa.

In Figure A8, we presented the raw $R\left(T, B_{appl}, P\right)$ data reported by Ying *et al.*[14], along with the deduced $T_{c,0.9}\left(B_{appl}, P\right)$ and $T_{c,0.1}\left(B_{appl}, P\right)$ data points for each resistive curve.

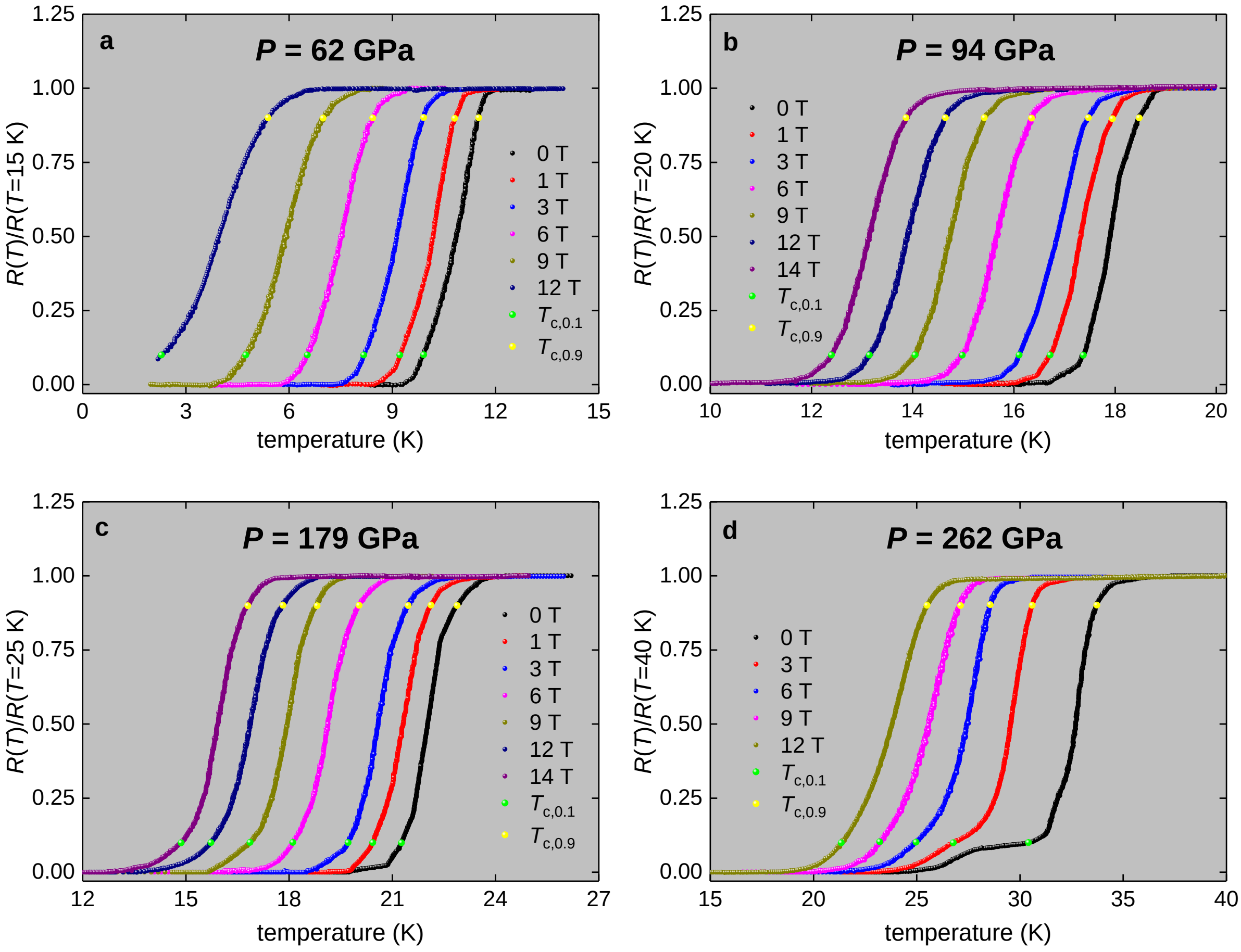

**Figure A8.** $R\left(T, B_{appl}\right)$ datasets and deduced $T_{c,0.9}\left(B_{appl}, P\right)$ (yellow circles) and $T_{c,0.1}\left(B_{appl}, P\right)$ (green circles) for highly compressed scandium. Raw $R\left(T, B_{appl}, P\right)$ curves were reported by Ying *et al.*[14]. (a) $P = 62\ GPa$; (b) $P = 94\ GPa$; (c) $P = 179\ GPa$;. (d) $P = 262\ GPa$.

In Figure A9, we presented the deduced $B_{c2}(T, P)$ values (defined by $\frac{R(T, B_{appl}, P)}{R(T_c^{onset}, B_{appl}, P)} = 0.10$ criterion) for the datasets showed in Figure A8. To deduce the ground state upper critical field values, $B_{c2}(0, P)$, we fitted the $B_{c2}(T, P)$ datasets to analytic approximant of the Werthamer-Helfand-Hohenberg[15,16] model, proposed by Baumgartner *et al.*[17]:

$$B_{c2}(T) = \frac{1}{0.693} \times \frac{\phi_0}{2\pi\xi^2(0)} \times \left(\left(1 - \frac{T}{T_c}\right) - 0.153 \times \left(1 - \frac{T}{T_c}\right)^2 - 0.152 \times \left(1 - \frac{T}{T_c}\right)^4\right) \qquad \text{(S2)}$$

where $\phi_0$ is the superconducting flux quantum, $\xi(0)$ is the ground state coherence length, and $\xi(0)$ and $T_c \equiv T_c\left(B_{appl} = 0\right)$ are free fitting parameters.

The derived ground state upper critical field values, $B_{c2}(0, P)$ (refer to Figure A9), were employed to compute the data for $\frac{\Delta T_c(B_{appl}, P)}{T_c(B_{appl}=0, P)}$ *vs.* $\frac{B_{appl}}{B_{c2}(0,P)}$ plots in Figure 4. Although the overall trend indicates broadening of the transition width with applied magnetic field, it´s noteworthy that each curve exhibits sections where the width does not increase.

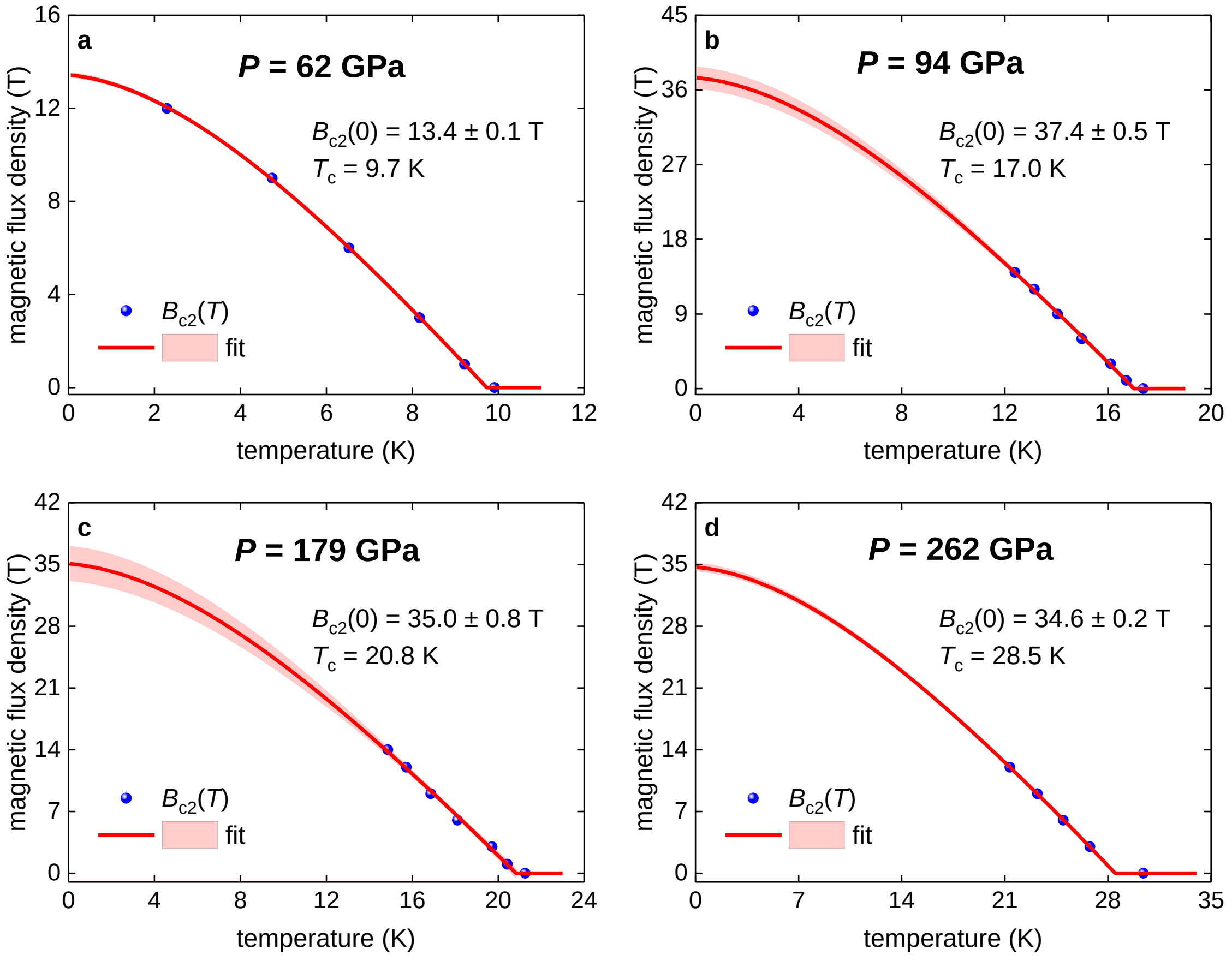

**Figure A9.** The upper critical field, $B_{c2}(T)$, deduced by the criterion of $T_{c,0.1}(B_{appl}, P) = \frac{R(T,B_{appl},P)}{R(T_c^{onset},B_{appl},P)} = 0.10$ for highly-compressed scandium and data fits to the model proposed by Baumgartner *et al.*[17]. The raw $R(T, B_{appl}, P)$ data were reported by Ying *et al.*[14]. (a) $P = 62\ GPa$; (b) $P = 94\ GPa$; (c) $P = 179\ GPa$;. (d) $P = 262\ GPa$. 95% confidence bands are shown by pink shadow areas.

In addition, we should stress that the definition of the resistive transition width by Eq. 1 is conventional, which implies that the transition width can be also defined by other conventions, and, for instance, by:

$$\frac{\Delta T_{c,0.95-0.05}(B_{appl},P)}{T_c(B_{appl}=0,P)} = \frac{T_{c,0.95}(B_{appl},P) - T_{c,0.05}(B_{appl},P)}{\left(\frac{T_{c,0.95}(B_{appl}=0,P) + T_{c,0.05}(B_{appl}=0,P)}{2}\right)} \qquad \text{(S3)}$$

where $T_{c,0.95}(B_{appl}) = \frac{R(T,B_{appl},P)}{R(T_c^{onset},B_{appl},P)} = 0.95$ and $T_{c,0.05}(B_{appl}) = \frac{R(T,B_{appl},P)}{R(T_c^{onset},B_{appl,P})} = 0.05$.

And if this transition width definition (Eq. S3) is applied to $R(T, B_{appl}, P = 262\ GPa)$ dataset (Fig. 4), one obtains the narrowing of the transition width, $\frac{\Delta T_{c,0.95-0.05}(B_{appl})}{T_c(B_{appl}=0)}$, for the field range of $\frac{B_{appl}}{B_{c2}(0)} \lesssim 0.15$, which is similar to ones deduced for $MgB_2$ (Figs. 3, A1,A2).

However, if the trend (i.e. the narrowing vs the broadening) depends on the definition of the transition width, $\frac{\Delta T_c(B_{appl})}{T_c(B_{appl}=0)}$, and there is no universal definition for the latter, we argue that the idea to use the $\frac{\Delta T_c(B_{appl})}{T_c(B_{appl}=0)}$ vs $\frac{B_{appl}}{B_{c2}(0)}$ trend as a criterion for the existence of the superconductivity in a new material is fundamentally incorrect and does not have any physical meaning.

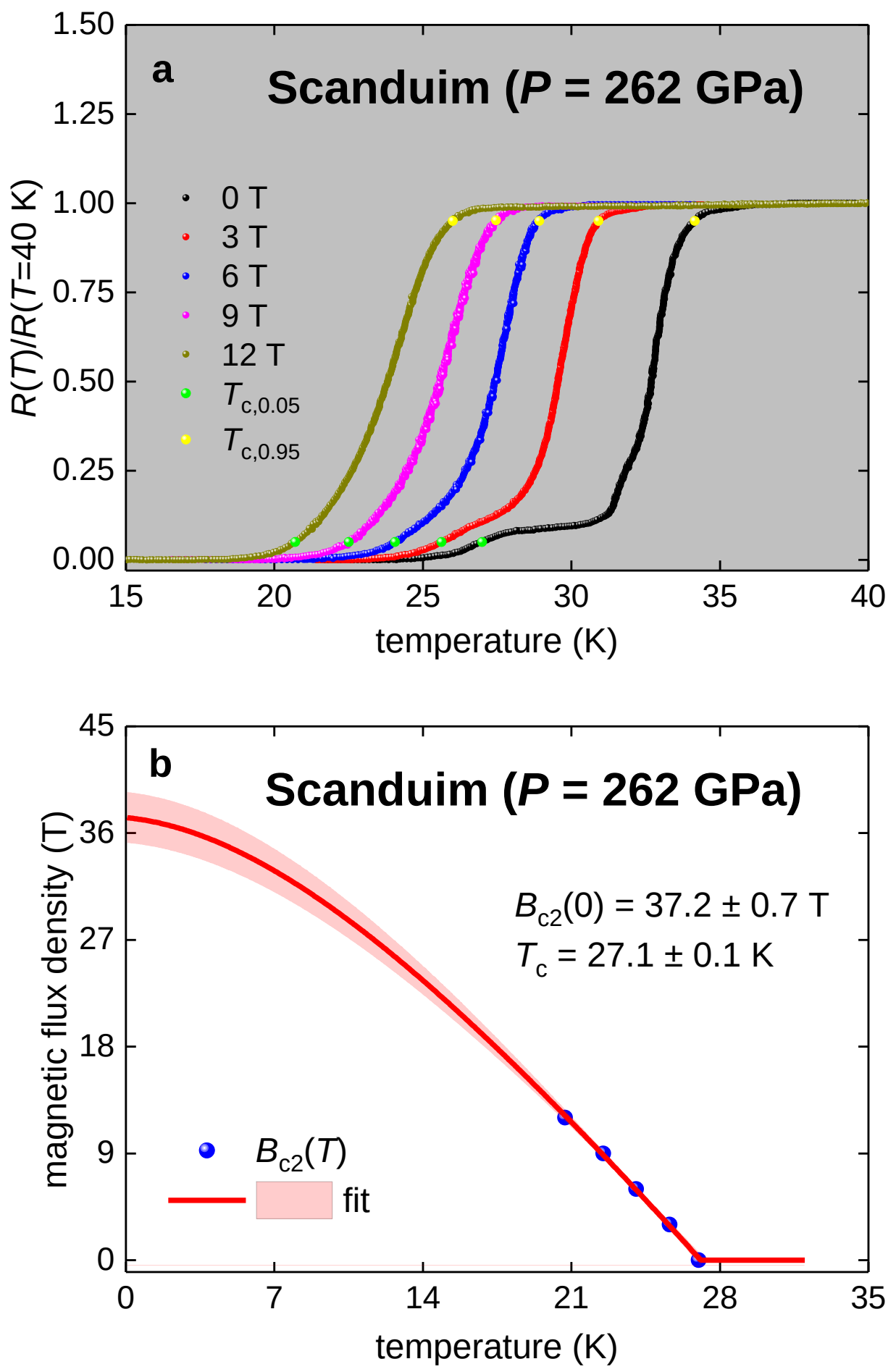

**Figure A10.** (**a**) $R(T, B_{appl}, P = 262\ GPa)$ dataset and deduced $T_{c,0.95}(B_{appl}, P)$ (yellow balls) and $T_{c,0.05}(B_{appl}, P)$ (green balls) for highly compressed scandium. Raw $R(T, B_{appl}, P)$ curves reported by Ying *et al* [S14]. (**b**) The upper critical field, $B_{c2}(T)$, deduced by the criterion of $T_{c,0.05}(B_{appl}, P = 262\ GPa) = \frac{R(T, B_{appl}, P=262\ GPa)}{R(T_c^{onset}, B_{appl}, P=262\ GPa)} = 0.05$, and data fit to model (Eq. S1) by Baumgartner *et al* [S8] for highly compressed scandium. 95% confidence bands are shown by pink shadow areas.

## Appendix E. Highly-compressed $H_3S$

Now we turn to hydrogen-based superconductors, and in Figure A11, we showed the $R(T, B_{appl}, P = 155\ GPa)$ data reported by Mozaffari *et al.*[18] for highly-compressed $H_3S$, where we highlighted $T_{c,0.1}(B_{appl}, P)$, $T_{c,0.9}(B_{appl}, P)$, and $T_{c,onset}(B_{appl}, P)$ data points.

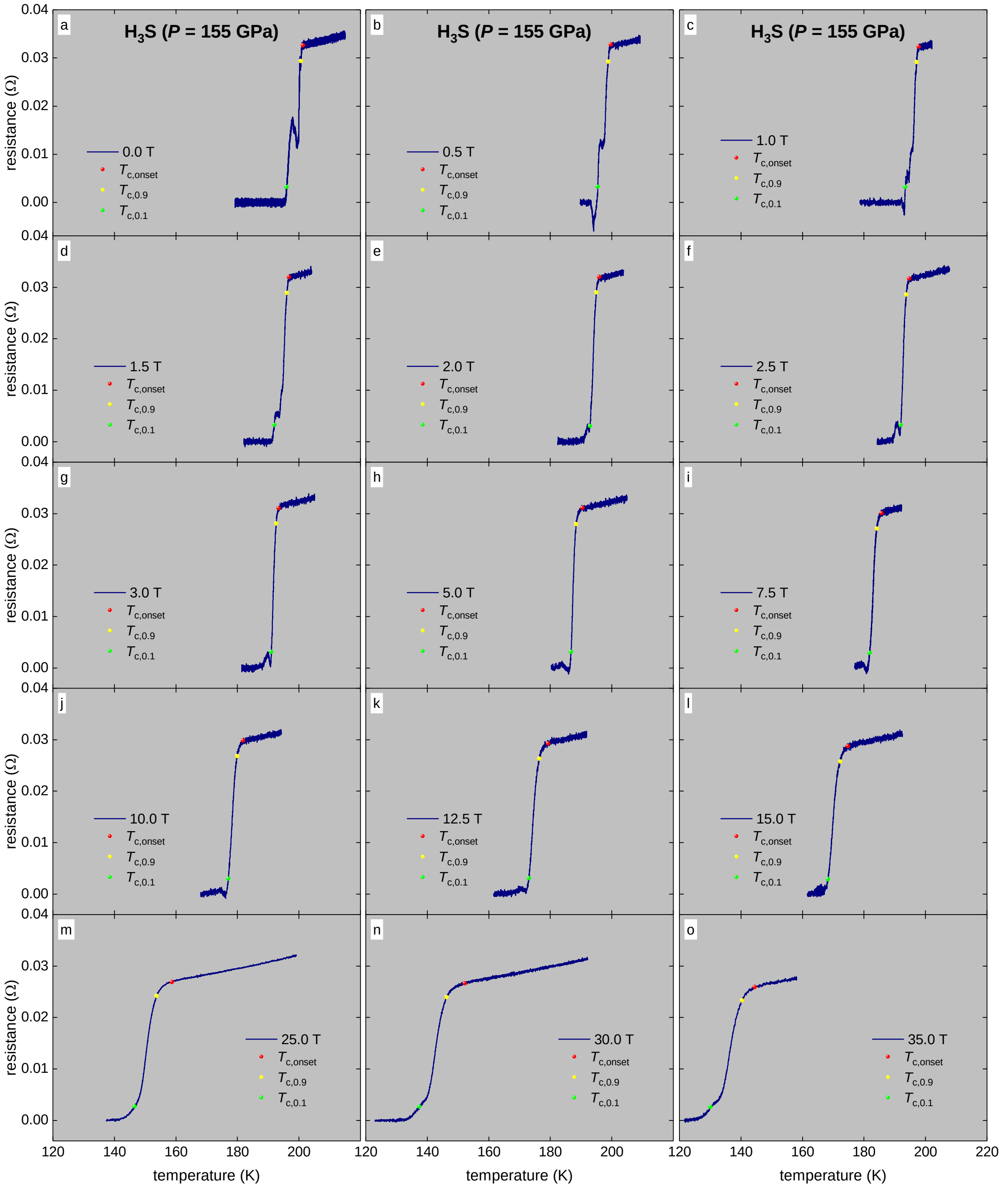

**Figure A11.** $R(T, B_{appl}, P = 155\ GPa)$ datasets for $H_3S$. Importantly, the raw $R(T, B_{appl}, P)$ data are freely available online[18]. Deduced $T_{c,onset}(B_{appl}, P)$ (red circles), $T_{c,0.9}(B_{appl}, P)$ (yellow circles) and $T_{c,0.1}(B_{appl}, P)$ (green circles) are marked for each $R(T, B_{appl}, P = 155\ GPa)$ curve.

To plot the $\frac{\Delta T_c(B_{appl}, P=155\ GPa)}{T_c(B_{appl}=0, P=155\ GPa)}$ vs. $\frac{B_{appl}}{B_{c2}(0, P=155\ GPa)}$ relationship for $H_3S$ (see Figure A12), we employed the reported value for the ground state upper critical field, $B_{c2}(0, P = 155\ GPa)$, which is 88 T[18].

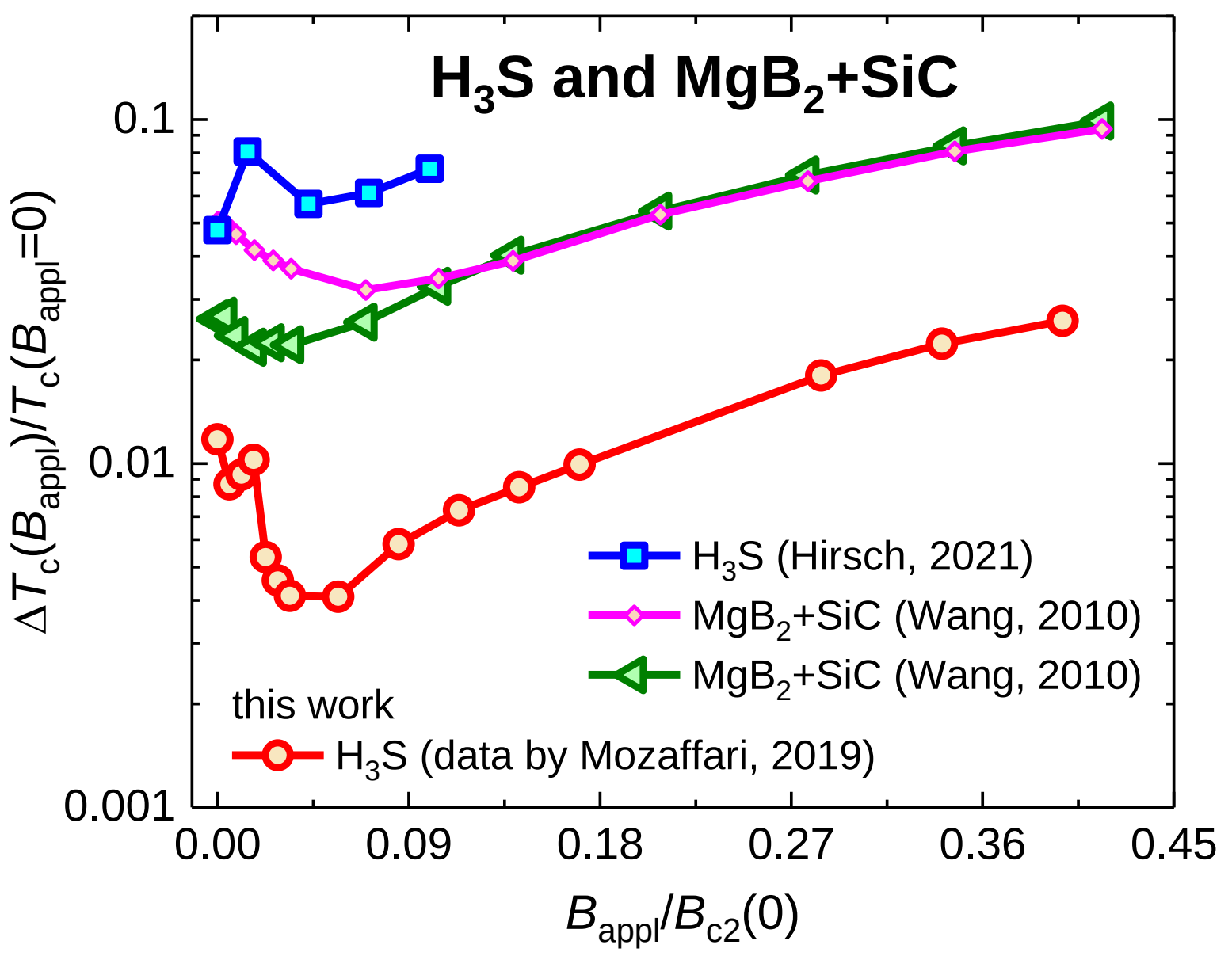

**Figure A12.** $\frac{\Delta T_c(B_{appl})}{T_c(B_{appl}=0)}$ *vs.* $\frac{B_{appl}}{B_{c2}(0)}$ relationship for highly compressed $H_3S$ ($P = 155\ GPa$): blue squares represent the data reported by Hirsch and Marsiglio[2], and red circles represent the data deduced in this work from the raw $R(T, B_{appl}, P)$ dataset reported by Mozaffari *et al.*[18]. Data for $MgB_2$ doped with SiC (green triangles and magenta diamonds) are from Wang *et al.*[6] (details are in Figures 3 and A2).

In general, we cannot confirm either the $\frac{\Delta T_c(B_{appl}, P=155\ GPa)}{T_c(B_{appl}=0, P=155\ GPa)}$ values nor the trend for the dependence reported by Hirsch and Marsiglio[2] for $H_3S$. It should be noted that these authors did not reveal the criterion for the $\frac{\Delta T_c(B_{appl}, P=155\ GPa)}{T_c(B_{appl}=0, P=155\ GPa)}$ definition and the method(s) used to extract the transition width from the raw datasets. Furthermore, the values reported by Hirsch and Marsiglio[2] are orders of magnitude larger than the values deduced in this work (see Figure A12). For instance, at $\frac{B_{appl}}{B_{c2}(0, P=155\ GPa)} \cong 0.05$, the difference between these values is 15-17 times greater.

To demonstrate that the $\frac{\Delta T_c(B_{appl}, P=155\ GPa)}{T_c(B_{appl}=0, P=155\ GPa)}$ *vs.* $\frac{B_{appl}}{B_{c2}(0, P=155\ GPa)}$ trend for $H_3S$ in this work aligns with the trend seen in conventional superconductors, we also included the $\frac{\Delta T_c(B_{appl})}{T_c(B_{appl}=0)}$ *vs.* $\frac{B_{appl}}{B_{c2}(0)}$ datasets for two samples of $MgB_2$ doped with SiC in Figure A12. The raw data were reported by Wang *et al.*[6], and additional details can be found in Figures 3 and A2.

## Appendix F. Highly-compressed $YH_6$ (P = 183 GPa)

The data on the observation of near-room temperature superconductivity in highly-compressed $YH_6$. $R(T, B_{appl}, P = 183\ GPa)$ by Troyan *et al.*[19], along with $T_{c,0.1}(B_{appl}, P)$, $T_{c,0.9}(B_{appl}, P)$, and $T_{c,onset}(B_{appl}, P)$, are shown in Figure A13.

This research group reported $B_{c2}(0, P = 183\ GPa) = 158\ T$, which we used to calculate the $\frac{\Delta T_c(B_{appl})}{T_c(B_{appl}=0)}$ *vs.* $\frac{B_{appl}}{B_{c2}(0)}$ relationship shown in Figure 4b.

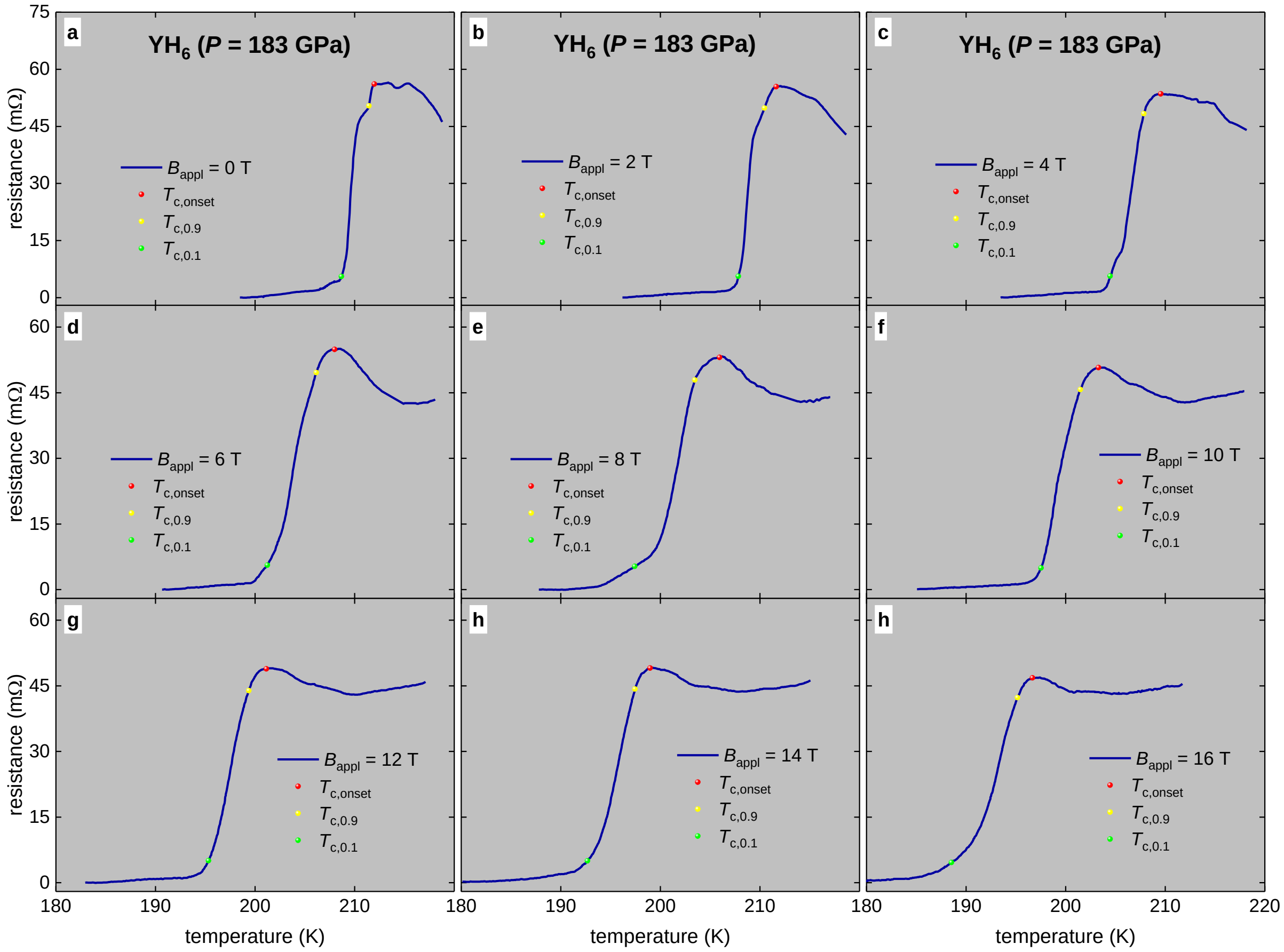

**Figure A13.** $R(T, B_{appl}, P = 183\ GPa)$ curves for the $YH_6$ phase (experimental data reported by Troyan *et al.*[19]). Deduced $T_{c,onset}(B_{appl}, P)$ (red circles), $T_{c,0.9}(B_{appl}, P)$ (yellow circles) and $T_{c,0.1}(B_{appl}, P)$ (green circles) for each $R(T, B_{appl}, P = 183\ GPa)$ curve are shown.

## Appendix G. Highly-compressed (La,Ce)$H_9$ (P = 110 GPa)

Bi *et al.*[20] reported on the observation of high-temperature superconductivity in highly-compressed (La,Ce)$H_9$. This research group reported $T_{c,0.1}(B_{appl}, P = 110\ GPa)$, $T_{c,0.9}(B_{appl}, P = 110\ GPa)$, and $B_{c2}(0, P = 110\ GPa) = 70\ T$, which we used to calculate the $\frac{\Delta T_c(B_{appl})}{T_c(B_{appl}=0)}$ *vs.* $\frac{B_{appl}}{B_{c2}(0)}$ relationship in Figure 4c.

## Appendix H. Highly-compressed $SnH_4$ (P = 180 GPa)

Troyan *et al.*[21] reported on the observation of high-temperature superconductivity in highly-compressed $SnH_4$. This research group reported $T_{c,0.1}(B_{appl}, P = 180\ GPa)$, $T_{c,0.9}(B_{appl}, P = 180\ GPa)$, and $B_{c2}(0, P = 180\ GPa) = 13.5\ T$, which we used to calculate the $\frac{\Delta T_c(B_{appl})}{T_c(B_{appl}=0)}$ *vs.* $\frac{B_{appl}}{B_{c2}(0)}$ relationship in Figure 4c.

## Appendix I. Highly-compressed hafnium hydride (P = 243 GPa)

Zhang *et al.*[22] reported on the high-temperature superconductivity in highly-compressed hafnium hydride. $R(T, B_{appl}, P = 243\ GPa)$ datasets reported by Zhang *et al.*[22] along with $T_{c,0.1}(B_{appl}, P)$, $T_{c,0.9}(B_{appl}, P)$, and $T_{c,onset}(B_{appl}, P)$ are shown in Figure A14(a).

The deduced $T_{c,0.1}(B_{appl}, P = 243)$ values were fitted to Eq. S2 (see Figure A14(b)), from which $B_{c2}(0) = 18.3\,T$ was derived. As a result, the monotonic increasing $\frac{\Delta T_c(B_{appl})}{T_c(B_{appl}=0)}$ *vs.* $\frac{B_{appl}}{B_{c2}(0)}$ relationship for highly compressed hafnium hydride was revealed in Figures 4c and A14(c).

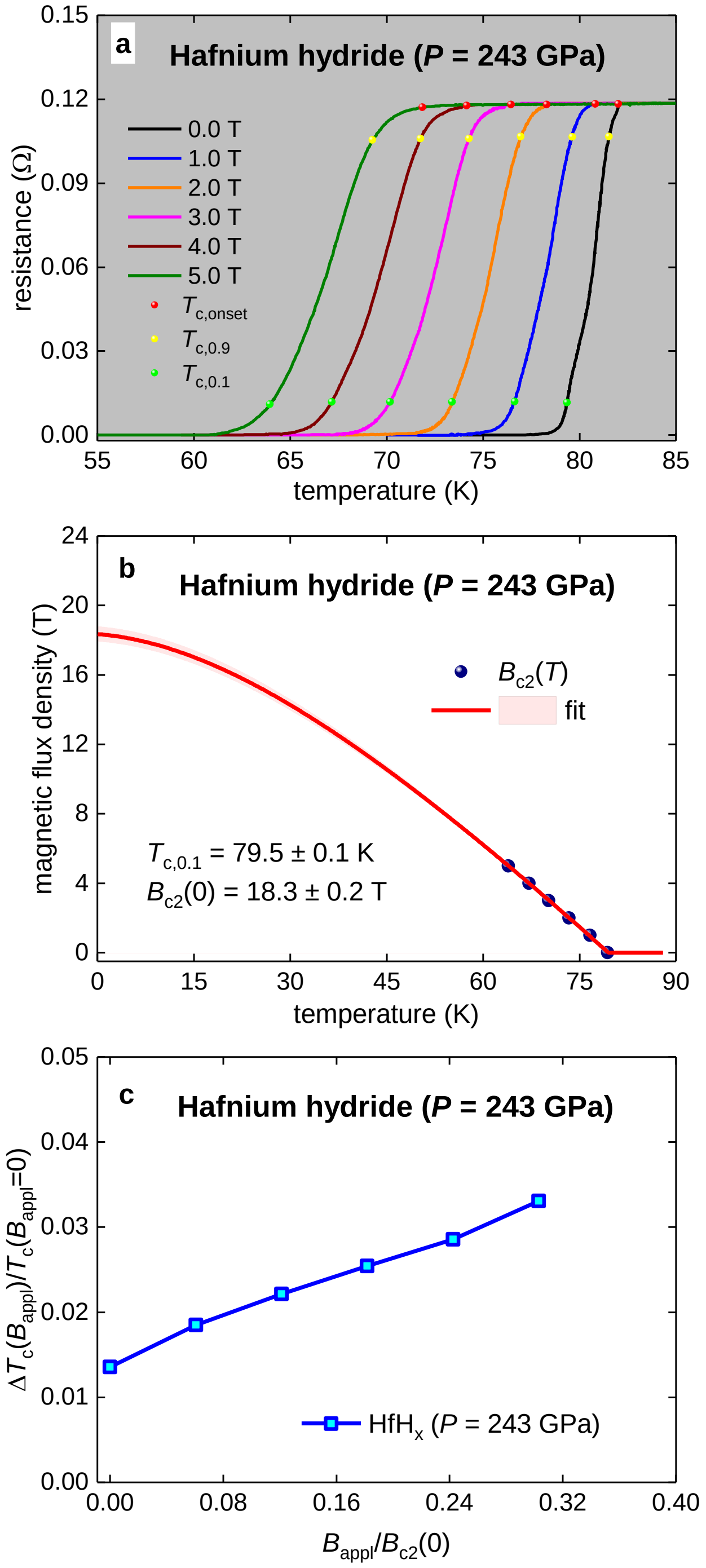

**Figure A14.** (a) $R(T, B_{appl}, P = 243\ GPa)$ datasets for $HfH_x$ (the raw $R(T, B_{appl}, P)$ data reported by Zhang *et al* [22]) with the deduced $T_{c,onset}(B_{appl}, P)$ (red circles), $T_{c,0.9}(B_{appl}, P)$ (yellow circles) and $T_{c,0.1}(B_{appl}, P)$ (green circles) values. (b) The upper critical field data, $B_{c2}(T)$, deduced by the criterion of $T_{c,0.1}(B_{appl}, P)$, and data fit to Equation S2. 95% confidence bands are shown by pink shadow areas. (c) The $\frac{\Delta T_c(B_{appl})}{T_c(B_{appl}=0)}$ *vs.* $\frac{B_{appl}}{B_{c2}(0)}$ relationship for $HfH_x$ $(P = 243\ GPa)$.

## Appendix J. Highly-compressed $CeH_9$ (P = 137 GPa)

Chen *et al.*[23] reported on the observation of high-temperature superconductivity in highly-compressed cerium hydride. The raw $R(T, B_{appl}, P = 137\ GPa)$ datasets, which are freely available online[23], along with $T_{c,0.1}(B_{appl}, P)$, $T_{c,0.9}(B_{appl}, P)$, and $T_{c,onset}(B_{appl}, P)$ values are shown in Figure A15.

The $\frac{\Delta T_c(B_{appl})}{T_c(B_{appl}=0)}$ vs. $\frac{B_{appl}}{B_{c2}(0)}$ relationship shown in Figure 4d was calculated by using $B_{c2}(0) = 22.9\ T$ reported by Chen *et al.*[23].

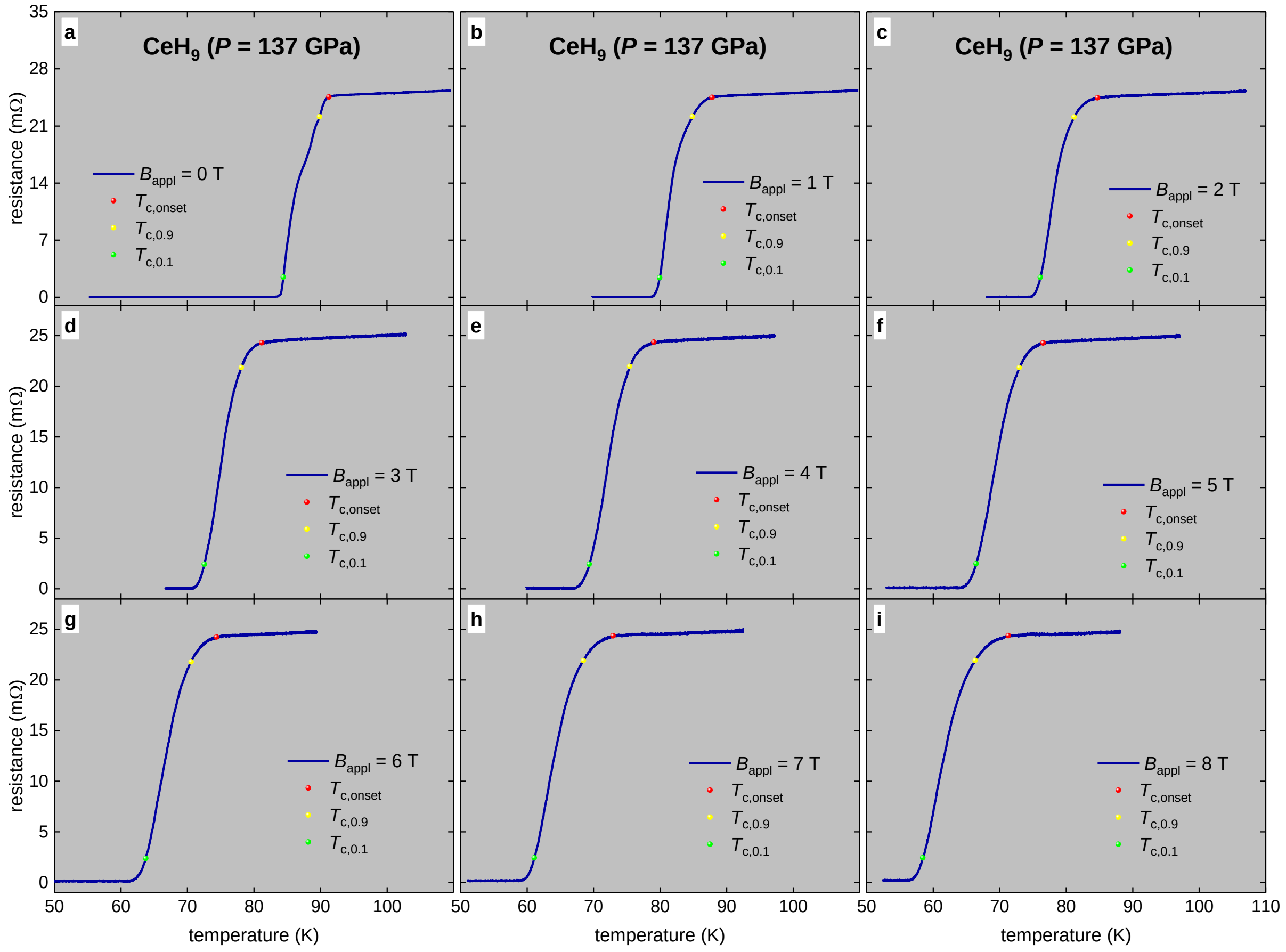

**Figure A15.** The raw $R(T, B_{appl}, P = 137\ GPa)$ datasets obtained during the cooling cycle of the $CeH_9$ phase. The deduced $T_{c,onset}(B_{appl}, P)$ (red circles), $T_{c,0.9}(B_{appl}, P)$ (yellow circles) and $T_{c,0.1}(B_{appl}, P)$ (green circles) values for each $R(T, B_{appl}, P = 137\ GPa)$ curve are shown.